\documentstyle[12pt,aasms4]{article}
\begin{document}

\title{The Evolution of Galaxies in Compact Groups}

\author{Roger Coziol\altaffilmark{1,2},
Reinaldo R. de
Carvalho\altaffilmark{2}}
\author{Hugo
V.Capelato\altaffilmark{3}}

\author{\and}

\author {Andr\'e L.  B. Ribeiro\altaffilmark{4}}

\altaffiltext{1}{Laborat\'orio Nacional de Astrof\'{\i}sica - LNA/CNPq, Rua Estados Unidos, 154, Bairros das Na\c{c}\~oes - 37500-000 - Itajub\'a, MG, Brasil}

\altaffiltext{2}{Observat\'orio Nacional, Rua Gal.  Jos\'e Cristino,
77 -- 20921-400, Rio de Janeiro, RJ., Brasil}

\altaffiltext{3}{Divis\~ao de Astrof\'{\i}sica -- INPE/MCT, C.P.  515
-- 12201-970, S.  Jos\'e dos Campos, SP., Brasil}

\altaffiltext{4}{Dept. Matem\'atica Aplicada - IMECC, Universidade E. de Campinas, 13083-970 SP, Brasil}

\begin{abstract}

We present the analysis of the spectra of 62 galaxies in 15 compact
groups. The galaxies were classified in four activity classes:
galaxies without emission, starburst galaxies, luminous AGNs (Seyfert
and LINERs) and low-luminosity AGNs (LLAGNs).

The star formation in the HCG starbursts is more intense than in normal
spirals, but comparable to those observed in starburst nucleus galaxies (SBNGs)
in the field. 
In general, the HCG starbursts have mean solar gas metallicity and do not
follow the metallicity--luminosity relation traced by
the early--type SBNGs in the field, suggesting that most of them are late--type SBNGs.  
This morphology preference coupled to the observation that the HCG starbursts are
predominantly in the halo of the groups is consistent
with the idea that compact groups are embedded in sparser structures.  

The stellar metallicities of the non starburst galaxies are
comparable to those observed in normal galaxies with similar morphologies,
but are relatively high for their luminosities.   
In these galaxies the metal absorption lines equivalent widths are slightly narrower 
than normal while the Balmer absorption lines are relatively strong. All these
observations suggest the presence of a population of
intermediate--age stars. These galaxies could be
``post--starburst'', but at a very advanced stage of evolution. 
The last bursts happened more than 2 Gyrs in the past.

Our observations are supporting a scenario where the core of the
groups are slowly collapsing evolved systems embedded in more extended
structures (Ribeiro et al. 1998).  In the core of the groups, the
interactions were more frequent and the galaxies evolved at a more
rapid rate than in their halos.

\end{abstract}


\section{Introduction}

One of the most important problems in cosmology concerns the
formation and evolution of galaxies.  
Recent works support the hierarchical galaxy formation theory, where massive galaxies
form by subsequent mergers of smaller mass of gas and star systems
(Larson 1990; Clements \& Couch 1996; Baugh et al.  1996; 
Rauch, Haehnelt \& Steinmetz 1997).  Although the details are still
largely missing, it seems that the starburst phenomenon could be an
important phase of this process (Sofue \& Habe 1992; Kennicutt
1994; Clements \& Couch 1996; Cowie et al. 1996; Simard \&
Pritchet 1997; Lowenthal et al.  1997; Coziol et al.  1997a).  In
the starburst--interaction scenario, tidal forces produced by
interaction are sufficient to trigger global bursts of star formation
resulting in changes of the stellar populations, chemical abundances
and even morphologies of the galaxies (Toomre \& Toomre 1972;
Tinsley \& Larson 1980; Kauffmann et al. 1996; Lavery et al.  1996;
Neuschaefer, Ratnatunga \& Griffiths  1997; Coziol 1997).  By considering various
mechanisms for transferring matter down to the nucleus of the
galaxies, one could extend the starburst--interaction scenario to
include the formation and evolution of AGNs (Weedman 1983;
Sanders et al. 1988; Carlberg 1990; Omont et al.  1996; Bahcall et
al. 1997). 

A natural consequence of the starburst--interaction scenario is that
the evolution of galaxies should depend on the environment in a
systematic way, being more effective in a dense, low--velocity dispersion region.  
One of the most important piece of evidence supporting the environment dependence
is the segregation of morphologies observed in galaxy clusters (Dressler
1980).  It is not clear however, how and when this distinction
occurs.  Although cluster ellipticals seem to be old and quietly 
evolving galaxies (Schade, Barrientos \& Lopez--Cruz 1997; Ellis et al.  1997; Bender et al. 1997), 
the history of the cluster spiral galaxies could be much different, experiencing 
more evolution until a recent epoch (Butcher \& Oemler
1978; Larson, Tinsley \& Caldwell 1980; Caldwell \& Rose 1997; Stein 1997;
Abraham et al.  1997; Dressler \& Smail 1997).

In this context of the formation and evolution of galaxies, Hickson's
Compact Groups (HCGs) of galaxies (Hickson 1982) are somehow
puzzling.  These groups combine high spatial densities and small
velocity dispersions suggesting rapid merging rates and possibly enhanced
star formation and nuclear activity (Hickson 1990; Rubin et al. 1991). Yet, 
the merging rates seem to be lower than that expected from the
observed crossing times (Zepf \& Whitmore 1991; Zepf 1993) and
the star formation and AGN activity are at a relatively low level (Coziol et al. 1997b, hereafter
Paper~1; 
Leon et al. 1998; Verdes--Montenegro et al. 1998). 
As an alternative, it was suggested that the HCGs are the result of chance alignments of
galaxies in larger structures (Mamon 1994; Ostriker et al.  1995).
But, this hypothesis seems to be ruled out based on the most recent
X-ray survey which suggests that as much as 75\% of these systems
present extended emission from hot intragroup gas (Ponmam et al.
1996).  

A more satisfactory answer to the ambiguities presented by
the HCGs comes from a new spectroscopic survey of galaxies performed
by de Carvalho et al.  (1997).  The dynamical analysis based on this
survey (Ribeiro et al. 1998; Zepf, de Carvalho \& Ribeiro 1997) reveals that most of
the groups are part of larger structures and represent
different dynamical stages of evolution.  Following the discovery of
a significant number of low--luminosity AGNs (LLAGNs) in HCGs, we verified
that the core of the groups are dominated by AGNs and galaxies
without emission, whereas the starbursts galaxies are mostly at the
periphery (Paper~1). From this result, we propose a
scenario where the type of activity (AGN or starburst) in galaxies correlates to a
morphology--density evolution.  The present contribution presents other
evidences in favor of this scenario and in support of the idea of a
replenishment of the groups by spiral galaxies during their
formation. If CGs are physically bound systems evolving on a timescale of
0.1 H$_{\circ}^{-1}$, then the first generation of these groups was
already converted into a field elliptical galaxy. Therefore, some replenishment
mechanism must operate in order to keep the number of groups approximately
constant (Diafero, Geller \& Ramella 1994, 1995;
Governato, Tozzy \& Cavaliere 1996; Ribeiro et al. 1998).

\section{Results of the spectral analysis}

In order to study the nature of compact groups of
galaxies, de Carvalho et al.  (1997) obtained the spectra of 316
galaxies in the regions of a selected sample of 17 HCGs, using the ARGUS fiber--fed
spectrograph at the 4m CTIO telescope.  The details on the
instrumental setup and a discussion of the data reduction can be
found in de Carvalho et al.  (1997).  From this sample,
we selected 82 galaxies which have a spectrum with a signal to noise
ratio sufficiently high to study their spectral characteristics. 
Of these galaxies, 28 (34\%) present only absorption lines and 54 (66\%)
present both emission and absorption lines.  A first paper (Coziol et
al. 1997b; hereafter Paper I) presented a classification of the activity types of the
emission--line galaxies. In the present contribution, we extend our analysis by
studying the absorption--line features of the galaxies 
which are group members. 

In Paper I, we have distinguished between the 
galaxies which are real members of the HCGs from those which are 
on the foreground or background. We note that many of 
these ``non-group members'' could reside
in other bound structures (Ribeiro et al. 1998).
Because our observations on these possible structures are very incomplete, 
we choose not to include them in the present study. 
Our final sample is composed of 62 galaxies which are 
members of 15 compact groups. 
 
In Table 1, we present the line intensities of the HCG emission--line galaxies.
The first column gives the identification of the galaxies
following de Carvalho et al.  (1997).  In columns 2 to 12, the
intensities are given relatively to H$\alpha = 100$.  The
uncertainties were determined from Poisson statistics.  Errors smaller
than 1\% are not indicated.  The lines were measured by fitting a
gaussian profile after the subtraction of a template galaxy spectrum.
This technique was applied systematically to
correct for the presence of strong Balmer absorption lines observed
in a significant fraction of the galaxies in our sample.  Details on
how this subtraction was performed can be found in Paper~I.

In Table 2, we summarize the results of our classification of the
activity type of the HCG emission--line galaxies as determined in Paper~I
and list some of their basic characteristics. The B magnitudes in
column 2 and the redshifts in column 3 were taken 
from de Carvalho et al. (1997). In column 4,
we list the absolute magnitudes of the galaxies, assuming H$_0 =
75$ km s$^{-1}$ Mpc$^{-1}$.  The morphological types listed 
in column 5 were taken from Mendes de Oliveira \& Hickson (1994).  The
different types of activity in column 6 are described as:  starburst
galaxies [SBNGs or HII], AGNs (Seyfert 2 [Sy2] or LINER [LNR]) and
low-luminosity AGNs [dSy2 or dLNR].  Column 7 gives the H$\alpha
+$ [\ion{N}{2}] equivalent width (EW).

The absorption features identified in the spectra of the galaxies 
were measured by drawing the pseudo continuum at the highest
points, using a region $\sim 100$ \AA\ wide on each side of the line
(Zabludoff et al. 1996).
This method was used because the spectra are not flux calibrated and
most of our galaxies are early-type galaxies with a
spectrum full of faint absorption lines. For the Balmer lines, we did not use a
narrow continuum window definition like in the Lick system (Faber et al.
1989) because this method clearly underestimate the equivalent width
of broad lines which are numerous in our spectra (Worthey et
al. 1994). We expect our method to yield slightly
overestimated EWs and peak intensities as compared to values
obtained using the Lick system. 

The determination of the Mg2 index follows the definition 
of Burstein et al. (1984), using their windows for the identification
of the pseudo continuum. In Jorgensen (1997) we have found 6 HCG galaxies
in common with our sample, where the Mg2 index was measured in the
Lick system. The values reported by Jorgensen show 
an offset of -0.11 as compared to ours. This difference probably\
reflects the fact that Jorgensen's measurements  were done with better
resolution. Thus, we have taken the offset of -0.11 to correct our Mg2
measurements. 

In Tables 3 to 6, we present the absorption line characteristics of 
the galaxies exhibiting different activity types. 
Columns 2 to 9 give the EW of the most prominent absorption features.  
The mean uncertainty of $\pm 0.5$ \AA\ was determined by comparing values 
obtained in two different spectra of the same object.  
Column 10 gives the ratio of the center of the line intensity of the \ion{Ca}{2} H +
H$\epsilon$ lines to the center of the line intensity of the \ion{Ca}{2} K 
and column 11 gives the Mg$_2$ index. Table 6 has an extra column which gives the
morphological type of the non--emission galaxies.

\section{The nature of the starburst in the HCGs}

In Figure 1, we compare the mean FWHM of the emission--lines identified in the 
spectra of the galaxies with their H$\alpha + $[\ion{N}{2}] EWs.  
On average, the AGNs (that is, the LLAGNs and the luminous AGNs) have 
smaller EWs and relatively broader FWHM than the starburst galaxies.  
The AGNs in our sample present FWHM which are within the range of typical 
observed values (Osterbrock \& Mathews 1986).  
Some of the starburst galaxies have unusually broad FWHM for \ion{H}{2}
regions, but comparable to those observed in Luminous Infrared
Galaxies (Veilleux et al. 1995).

Because our spectra are not flux calibrated we cannot determine directly
the star formation rates in the HCG starbursts. However, we can get a qualitative estimate 
of the level of star formation activity in these galaxies 
by examining their H$\alpha + $[\ion{N}{2}] EWs. 
It is well known, in particular, that galaxies with very
high equivalent width ($\ge 50 $\AA\ ) invariably prove
to be galaxies undergoing intense star formation
(Kennicutt \& Kent 1983). In Figure 2, we compare the distribution
of the H$\alpha + $[\ion{N}{2}] EWs of the HCG starbursts to those observed
in normal spirals (Kennicutt 1983). 
The HCG starbursts usually have relatively high EWs values, suggesting
that the star formation in these galaxies is on average more intense than 
in normal spiral galaxies. This observation is important because it
means that we cannot relate the star formation activity in these
galaxies to normal star formation occurring in the disk of late--type
galaxies. This alternative was recently proposed by Rakos, Maindl \& Schombert (1996) to
explain the Butcher--Oemler effect in rich clusters of galaxies at higher redshifts. 
In the HCGs starbursts, the star formation rates seem higher than normal, although
comparable to those observed in starburst galaxies in the field. 

Table 7 gives the gas densities and metallicities of the starburst
galaxies in our sample.  Column 2 shows the ratio
[\ion{S}{2}]$\lambda6717$/[\ion{S}{2}]$\lambda6731$ from which we
deduce the electron densities, in column 3, assuming a gas temperature of 10$^4$ K
(Osterbrock 1989). The densities are comparable to those found in 
giant \ion{H}{2} regions in the nucleus of galaxies. The metallicities, in column 4, were determined
based on the metallicity--calibrated diagnostic diagram established
by Coziol et al. (1994) and presented in Figure 3.  The mean uncertainty on the
metallicities is 0.2 dex.
In Figure 3, the metallicities of the HCG starbursts are compared to those
of a sample of Starburst Nucleus Galaxies (SBNGs) observed in the field 
(Coziol et al. 1997a, 1997c). 
As it can be seen, the mean metallicities of the HCG starbursts 
are comparable to those of the SBNGs in the field. 

In Coziol et al. (1997a) it was shown that in the field
the late--type SBNGs are 0.2 dex more metal rich on average than the early--type
SBNGs. This difference in metallicity was
explained by assuming different gas accretion rates during the formation
of the galaxies: the late--type SBNGs have accreted more gas than stars
than the early--type SBNGs (Coziol et al. 1997a).
In Paper I, we noted that most of the HCG starbursts in our sample 
with a well-defined morphology are late--type galaxies.
In Figure 3, we see that the mean metallicity of the HCG starbursts is
close to solar. The HCG starbursts look, consequently, more similar to 
the late than to the early--type SBNGs. This is another argument suggesting
that most of the HCG starbursts in our sample are in late--type spiral galaxies.
To test this hypothesis, we have estimated the difference ([O/H]$_{th}-$[O/H]$_{obs}$)
between the metallicity predicted by the luminosity--metallicity
relation followed by the early--type SBNGs (Coziol et al. 1997a) 
and the observed metallicity of the HCG starbursts. 
In Figure 4, we can see that the metallicities of the HCG starbursts generally diverge from
the values predicted by the luminosity--metallicity relation.
The distribution of the differences is similar to 
the one observed for the late--type SBNGs in the field. 
A Kolmogorov--Smirnov test indicates that there is no difference between
the Late--type SBNGs and HCG starburst distributions at the 5\% significance level. 
Therefore, unless they are completely different from those in the field, 
most of the HCG starbursts in our sample must be late--type SBNGs.

\section{The nature of the stellar population in the HCG galaxies}

The study of the absorption lines in galaxies
allows to determine the nature of the dominant stellar populations. 
By comparing the stellar populations in the HCG galaxies with
those found in normal galaxies in the field we can verify
if the evolution of these galaxies was influenced by their environment. 
But before doing such a comparison, we have to establish what is the 
observed pattern for normal galaxies, starburst and AGN galaxies in the field. 
 
For normal galaxy spectra, 
the absorption features vary with their morphologies.  
This is due to a variation of the nature of the dominant
stellar population in galaxies with different morphologies (Hamilton
1985; Rose 1985; Kennicutt 1992; Zaritsky, Zabludoff \& Willick 1996). 
In general, the EWs of different absorption lines 
trace a continuous sequence which correlates to morphology.
This phenomenon is illustrated in Figure 5a where we show
the behavior for the EW of the \ion{Ca}{2} K line 
as a function of the EW of the G-band in
a sample of normal spiral galaxies observed by Kennicutt (1992). The EWs were
measured here by Jablonka \& Alloin (1995). 
The EWs increase from the late--type to the early--type galaxies  
and follow the increases of age of the dominant stellar population.

In normal spiral galaxies, the dominant stellar population changes  
from the disk to the bulge. The stellar populations 
are generally older in bulges than in disks. This is illustrated
in Figure 5b, where we show the behavior of 
the EW of the \ion{Ca}{2} K line and of the G-band
lines, as measured in the bulges of spiral galaxies by Jablonka et al.
(1996). The EW of these two lines are comparable to those observed 
in elliptical galaxies (Jablonka et al. 1996).
This result is important as it suggests that the bulk of the stars in the bulges 
probably formed by similar processes than in elliptical 
galaxies (Jablonka et al. 1996). 

In the starburst and AGN galaxies, the change in the absorption line features 
with the morphologies are different than in the normal spiral galaxies.
In Figure 5b, we show the relation of the EW of the \ion{Ca}{2} K line with
the one for the G-band, as
measured in a sample of starburst galaxies by Storchi--Bergmann, Calzetti \& Kinney (1994).
The low values of the EWs suggest that the spectra of the starbursts
are dominated by young stellar populations, regardless of their
morphologies. Therefore, we cannot determine the morphology of a starburst galaxy 
by the absorption features in its spectrum. 

For the AGNs, the situation is more ambiguous than for the starbursts.
In Figure 5c, we show the EW of the \ion{Ca}{2} K line and of the G-band, as
measured in a sample of Seyfert galaxies by Storchi--Bergmann, Kinney \& Challis (1995).  
In general, the Seyfert galaxies seem to follow the same morphological sequence 
as normal galaxies, but the dispersion in each morphological bin is large. 
In this kind of analysis we should consider a possible dilution
effect due to the underlying featureless continuum of the AGN. 
Recently Cid Fernandes, Storchi-Bergman \& Schmitt (1998) have shown
that in the optical a dilution effect is observed only in Seyfert 1,   
not in Seyfert 2. In this latter case, therefore, the high dispersion  
of EWs values observed probably reflects the fact Seyferts 2 do
not constitute, from the point of view of star formation,
a homogeneous class of galaxies: some of these 
galaxies could be starburst while others are normal. This interpretation
is supported by the recent observations showing intrinsic differences
between the two types of Seyferts 
(see Gonz\'alez Delgado et al. (1997) and Malkan, Gorjian \& Tam (1998)).  
The consequence for our analysis is that when an AGN 
is present the EWs of the galaxies are not clearly related to their morphologies.  

From the previous discussion, we discriminate in Figure 5d
three different regions related to three dominant stellar populations
in galaxies: early--type galaxies, [E], which are dominated by an old
stellar population; intermediate and late--type spirals, [L], which
are dominated by intermediate--age and young stellar populations; and
starburst galaxies, [SB], which are dominated by young and massive
stars, independent of their morphologies.  Having determined these
regions, we can proceed to compare the stellar
population in our sample of HCG galaxies with those of normal galaxies in the field.  
Figure 5d shows the results for the HCG galaxies with different activity
types. The behaviors of the HCG starbursts and AGNs
are consistent with the general trend observed for galaxies of their type.
On the other hand, the HCG galaxies without emission and the LLAGNs all have 
smaller EWs than normal galaxies for their morphologies. This is surprising, 
because most of these galaxies are early--type and, consequently,  we would have
expected to find them clearly located in the E box. 

In order to verify that this phenomenon is not due to spurious measurements
of some absorption lines (the \ion{Ca}{2} K line, for instance, is
very near the border of our spectra and consequently hard to
measure), we show in Figure 6 the same kind of analysis using the CN
band.  The behavior is the same:  the non--emission galaxies
and the LLAGNs have smaller EWs than normal galaxies for their morphologies. Note
that following our methodology we expect to overestimate the EW values, which means
that the real effect could even be more pronounced than what we have measured.

In Figures 5 and 6, there seem to be no
difference between the stellar populations of the non--emission line
galaxies and the LLAGNs. It suggests that the AGN continuum in the
LLAGNs has no effect on the EWs absorption lines due to their
stellar population (the continuum of the AGN is probably too weak
as already pointed out in Paper~I).   
It seems, therefore, that the only way to explain the small 
EW values in all these galaxies is to assume the presence of 
younger stellar populations than usually observed in galaxies 
with similar morphologies.  

To test this hypothesis, we examine in Figure 7 the
behavior of the EW of H$\delta$ as a function of the ratio
(\ion{Ca}{2} H + H$\epsilon$)/\ion{Ca}{2} K. The increase of the EW of
the Balmer lines due to the presence of young stellar populations 
is already well established and documented (Dressler \& Shectman 1994; Leonardi \&
Rose 1996; Caldwell et al.  1996; Zabludoff et al.  1996; Poggianti
\& Barbaro 1996; Barbaro \& Poggianti 1997). The behavior of the (\ion{Ca}{2} H +
H$\epsilon$)/\ion{Ca}{2} K ratio due to the evolution of stellar populations is
explained by Rose (1985). This parameter was used recently by
Leonardi \& Rose (1996) to estimate the age of  
post--starburst galaxies in Coma.  In Figure 7, we can see that
the majority of the non--emission and LLAGN galaxies in our sample
have higher values of EWs than normal galaxies for their
morphologies. A mean value of EW(H$\delta$) $= 1.7$ is considered normal for 
late type spirals while EW(H$\delta$)$ = 2.5$
is considered unusually ``strong'' (Poggianti \& Barbaro 1996, 1997; 
Zabludoff et al. 1996).  

Note that the EW(H$\delta$) values found for HCG post--starburst galaxies seem relatively low
in comparison with those measured in post--starburst galaxies found
in the literature. This suggests that the ``post--starburst'' phase in the HCG galaxies is at 
a very advanced stage of evolution. This interpretation is consistent
with the observed ratios (\ion{Ca}{2} H + H$\epsilon$)/\ion{Ca}{2} K, which 
indicate a fraction of A type stars lower than 30\% (Rose 1985). 
Using the burst models of Poggianti \& Barbaro (1996) 
and Barbaro \& Poggianti 1997, a rough estimate would place the last star 
formation burst more than 2 Gyrs in the past.   

\section{Stellar metallicities of the galaxies in the compact groups}

In this section, we compare the stellar metallicities of 
HCG galaxies with those observed in normal galaxies  
(that is, non starburst and non AGN galaxies) with different morphologies.
Our analysis is based on the Mg$_2$ index which is generally used
to determine the metallicity of the stellar population
in composite systems (Burstein et al. 1984; Brodie \& Huchra 1990; Worthey, Faber \& Gonz\'alez 1992).

In Figure 8a, we show the distribution of the Mg$_2$ index as a
function of the morphological type in normal galaxies.  The normal
galaxies population is represented by elliptical and
lenticular galaxies observed by Bender et al. (1993) and by a sample of
spiral and lenticular bulges of galaxies observed by Jablonka et al. (1996).  
As expected, the normal galaxies trace a sequence of increasing 
metallicities, going from the late--type spirals to the
early--type galaxies. A similar sequence is observed for the
metallicity of the gas in the spiral galaxies (Zaritsky et al. 1996). 

The Mg$_2$ index for the HCG galaxies are presented in
Figure 8b. In general, the
stellar metallicities in the HCG starbursts are comparable to those observed
in the bulge of the late--type spiral galaxies. This reinforces our thesis 
that the HCG starbursts are mostly late--type spirals.  
The stellar metallicities of all the other non starburst HCG galaxies are 
also comparable to those observed in galaxies with similar morphologies. 
Note however that the average metallicity is high. In particular, the late--type HCG
galaxies seem to have metallicities comparable to those of the early--type.  
The tendency of high metallicities for the late-type HCG galaxies
in the HCG galaxies is better seen in Figure 9, where
we present the Mg$_2$ indices as a function of the absolute magnitudes of the galaxies.
On average, the Mg$_2$ index is 0.04 mag higher for the  HCG non starburst galaxies
than for early--type galaxies with similar luminosities. 
In Figure 9, we have also reported the values of Mg$_2$ as measured by Jorgensen (1997) which
show the same effect. 
Using the relation [Fe/H] $ = 9.92\ \times$ Mg$_2 - 2.21$, as proposed
by Brodie \& Huchra (1990), we found an excess of 0.4 dex in metallicities. 

\section{Discussion}

The main results of our analysis of the emission and absorption
features of the galaxies in the compact groups may be summarized 
as follows:

\begin{enumerate} 

\item Most of the HCG starbursts are late--type spirals.  The
star formation in these galaxies is higher than in normal spirals,
but comparable to those found in starburst galaxies in the field. Compared to 
the other types of galaxies in the groups, the HCG starbursts have slightly
lower metallicities, suggesting that their evolution was somehow different
compared to the other galaxies in the group. 

\item The non--emission line galaxies and the
LLAGNs show unusually small equivalent widths and
strong balmer lines for their morphologies. This suggests the presence of
intermediate--age stellar population. However, the fraction of these stars
seem to be much lower than in post--starburst galaxies found in the literature, suggesting 
a more advanced stage of evolution. 

\item All the non starburst HCG galaxies have Mg$_2$ indices which are
higher than for normal galaxies for their luminosities. In terms of  
metallicity ([Fe/H]), this would correspond to an excess of
almost 0.4 dex. 
 
\end{enumerate}

Our observations are consistent with a scenario where the formation
and evolution of galaxies in the compact groups are regulated by their environment.  
Merger, starbursts and/or gas stripping are probably 
the main processes behind the formation and evolution of these systems.
Evidence of gas stripping is inferred from the fact that HCG galaxies are
generally deficient in HI (Huchtmeier 1997; OOsterloo \& Iovino 1997). 
This process seems also a reasonable mechanism to explain the presence
of diffuse X-ray emission observed in the groups (Ponman et al. 1996).
The gas stripping mechanism would also produce false post--starburst galaxies
by truncation of star formation: the constant
star formation of a spiral is artificially stopped due to the 
rapid depletion of the gas favoring the detection of the intermediate age
stellar population. Skillman et al. (1996) have also
suggested that the curtailment of the metal poor gas infall produced by gas stripping
would contribute to increase substantially the gas metallicity of these galaxies. 
Note however, that in the HCG galaxies it is the stars which seem to 
be more metal rich. This implies that these stars were formed from previous
star formation events, or that these galaxies are real post--starburst
(similar phenomena is observed in Coma, see Caldwell \& Rose 1997).
This is consistent with the observations 
by Verdes--Montenegro et al. (1997) which show CO depletion
in the HCG galaxies. As noted by these authors, gas 
exhaustion by tidally induced star formation is probably the only way to produce the
depletion of molecular gas, since tidal stripping is less likely for molecular gas concentrated 
in the inner disks. Other elements in favor of past induced star formation
either by merger or interactions are the observations, also by Verdes--Montenegro et al. (1997),
of the presence of molecular gas in early--type galaxies (like in HCG 90 and HCG 86) and also
the fact that the galaxies deficient in molecular gas are located in the most compact, and 
consequently dynamically most evolved groups.    

But the strongest argument in favor of the role of merger  
for galaxy evolution in groups is the
density--morphology--activity trend observed in HCGs (see Paper~I):
the most massive and early--type galaxies are AGNs or non--emission
galaxies, mainly located in the core of the groups. 
In Ribeiro et al. (1998), it was noted that most of the HCGs are embedded
in larger structures forming extended halos around more dense and dynamically 
distinct cores displaying low velocity dispersions.  
Figure 10 shows the frequency distribution of peculiar velocities of the
HCG galaxies of our sample relative to their group mean
velocity. There is a clear tendency for the AGNs
and galaxies without emission to have smaller peculiar velocities than the starbursts.  
Now, not only these galaxies have early-type morphologies
but they are also more metal rich than the starburst galaxies which are
predominantly in the halos. Therefore, the {\it the cores are in a more
advanced stage of evolution than the halos}. In the cores, because of
their high density and low peculiar velocities, the interactions are
more frequent and the galaxies evolve  more rapidly
than in the halos of the groups. 

The fact that the galaxies in the core 
of the groups are more evolved than those in the halo only exacerbate 
the low crossing time paradox related to the evolution of these structures.   
Indeed, our observations suggest that the last burst of 
star formation occurred more than 2 Gyr in the past and that the present 
rates of star formation in these galaxies are relatively low. These
observations are inconsistent with the short merging and dynamical time scales. Numerical
simulations indicate that these systems are dinamically unstable, with an
orbital decay timescale between 0.01 and 0.1 H$_{\circ}^{-1}$ (Barnes 1985, 1989).
One solution, other than assuming the presence of a large amount 
of dark matter, would be to suppose that the core of the groups
formed relatively recently from the merging of smaller mass galaxies.
One example of this process still going on today would be 
HCG 16 (Ribeiro et al. 1996, Coziol \& de Carvalho 1998, in preparation).   
 
In the halos of the groups, the
interactions are less frequent due to the lower densities and the
higher peculiar velocities of the galaxies, allowing the formation and survival of
the spirals over longer time scales. In this sense the fact that all HCG starbursts may be
late--type spirals could be {\it very} significant.

Our observations are consistent with a scenario where the core of the groups are
slowly collapsing and evolved structures embedded in a more extended system (Ribeiro
et al.  1998).  The fact that spirals in the halo are starbursts
suggests that there is still a lot of activity going on in the halo.
Following this activity, some of the starburst galaxies in the halo
may fall into the core and enrich it in spirals.   

\acknowledgments

We would like to thank the referee Dr. Gregory Bothun for 
his remarks and suggestions which greatly contributed to improve the quality of
this paper. We also thank Dr. Eduardo Telles for stimulating discussions on the nature
of the starburst galaxies. R. Coziol acknowledges the financial support of the CNPq, under
contracts 650018/97-4 and A. L. B. Ribeiro acknowledges the support of
the Brazilian CAPES.

\clearpage 
\figcaption[coziolII_fig1.eps]{Mean FWHM for the
emission--line galaxies as a function of the EW(H$\alpha +$ [NII]).
The domain of values for the FWHM usually observed in the narrow line
regions of AGNs is indicated. The FWHM for
the starbursts are unusually broad, but comparable to those observed
in the Luminous Infrared Galaxies. In starbursts, the EW(H$\alpha +$
[NII]) is directly related to the star formation rates, which increase in the sense
indicated by the arrow. Galaxies with log(EW(H$\alpha +$
[NII])$ > 1.7$ are undergoing intense star formation (Kennicutt \& Kent 1983).}

\figcaption[coziolII_fig1.eps]{Frequency distribution of the EW(H$\alpha +$
[NII]) of the HCG starbursts. The HCG starburst are compared to the early and late--type spirals from 
the sample of Kennicutt \& Kent (1983). The mean and dispersion values are, $84\pm71$
for the HCG starbursts, $6\pm7$ for the early--type spirals and $33\pm20$ for the 
late--type spirals. 85\% of the HCG starbursts have EWs higher than 20 and 70\% have EWs
higher than 40.}

\figcaption[coziolII_fig3.eps]{The diagnostic diagram of line ratios
calibrated in metallicity (Coziol et al.  1994). The numbers
correspond to different values of metallicities, as given by 12 +
Log(O/H) (solar metallicity is 8.9 in this scale).  
The mean metallicities and dispersion for early--type and late--type 
SBNGs in the field are reported for comparison. 
The mean metallicity of the HCG starbursts is similar
to those of the late--type SBNGs in the field.}

\figcaption[coziolII_fig4.eps]{ The difference between the metallicities observed 
and the values predicted by the luminosity--metallicity relation
followed by the early--type SBNGs in the field. In general, the HCG starbursts
have metallicities that diverge from the luminosity--metallicity relation.
The distribution of the differences is similar to the one for
the late--type SBNGs in the field. }

\figcaption[coziolII_fig5.eps]{EW-EW diagram of the G--band as a
function of the CaII K lines:  a) Normal spirals show a sequence of
increasing EW towards the early--type morphologies (Jablonka \&
Alloin 1995); b) The bulge of spirals are dominated by an old stellar
population (Jablonka, Martin \& Arimoto 1996), while the starbursts [SB] are
dominated by young and massive stars, irrespective of their
morphologies; c) From the stellar population viewpoint, the Seyfert
galaxies do not form a homogeneous sample; d) We identify 3 regions
related to 3 different dominant populations:  [E] old population, [L]
intermediate--age and young population, [SB] young and massive stars.
In the HCGs, the LLAGNs and the galaxies without emission have
smaller values of EW than normal galaxies for their morphologies,
suggesting the presence of intermediate--age stellar population.}

\figcaption[coziolII_fig6.eps]{EW--EW diagram for the G--band as a
function of the CN line.  The same conclusion as in Figure 4d is
reached for the HCG galaxies.}

\figcaption[coziolII_fig7.eps]{The EW(H$\delta$) as a function of the
ratio of the intensities of lines CaII H + H$\epsilon$ over CaII K.
Typical values for normal galaxies of different morphologies
are indicated (Barbaro \& Poggianti 1997). 
The two vertical lines indicate the ratios CaII H + H$\epsilon$
for an old stellar population with 30\% and no A stars present (Rose 1985). 
All the HCG galaxies have strong Balmer EW and relatively small CaII H +
H$\epsilon$/CaII K ratio, consistent with the evolved post--starbursts hypothesis.}

\figcaption[coziolII_fig8.eps]{The Mg$_2$ index as a stellar
metallicity indicator.  a) The stellar population in normal galaxies
trace a sequence of metallicities, which increase towards the
early--type galaxies; b) The stellar metallicities of the HCG
galaxies are compared to those of normal galaxies; the mean
metallicities and the range values for normal early (T$\leq 0$) and
late--type galaxies are represented by two boxes. The HCG
galaxies without morphological information were placed at the two extremes
of the figure. Typical uncertainty is indicated as error bar at the
right of the figure.}

\figcaption[coziolII_fig9.eps]{The Mg$_2$ index in function of the
luminosities. The HCG galaxies are compared with the early--type galaxies 
from Bender, Burstein \& Faber (1989; BBF) and the bulge of spirals from Jablonka, Martin \&
Arimoto (1996; JMA). The solid line is a linear fit on the 
Bender, Burstein \& Faber data. The dashed line is the same fit offset by +0.04 mag.
We also show the error bar on our measures. The values of Mg$_2$ of 6 HCG galaxies
as observed by Jorgensen (1997) are also shown for comparison.}

\figcaption[coziolII_fig10.eps]{Frequency distribution of the peculiar 
velocities of the
emission--line galaxies relative to their group mean
velocity. The AGNs and the galaxies without emission show a clear tendency
to have smaller peculiar velocities than the starbursts.}

\clearpage
\begin{deluxetable}{lccccccccccc}
\footnotesize
\tablecaption{Spectral line ratios relative to H$\alpha$\label{tbl-1}}
\tablewidth{0pt}
\tablehead{
\colhead{HCG  \#}       &\colhead{$\lambda$4101} &\colhead{$\lambda$4340} &\colhead{$\lambda$4363} &
\colhead{$\lambda$4471} &\colhead{$\lambda$4861} &\colhead{$\lambda$5007} &\colhead{$\lambda$5876} &   
\colhead{$\lambda$6300} &\colhead{$\lambda$6584} &\colhead{$\lambda$6717} &\colhead{$\lambda$6731}
}
\startdata
04  01 &0.84   &   4.25  & \nodata& \nodata&   21.19&    2.04&  3.63  &  1.17   &   49.05&   8.87  &  6.92  \nl
04  03 &\nodata&  \nodata& \nodata& \nodata&   17.75&    7.60& \nodata&  2.44   &   47.48&  12.29  &  8.68  \nl
04  11 &5.65   &  13.48  & 2.87   & 2.82   &   46.56&  271.77&  6.86  &  1.10   &    5.43&   4.50  &  4.70  \nl
16  01 &\nodata&  \nodata& \nodata& \nodata&    7.8 &   16.0 & \nodata& 13.5    &  123.5 &  29.9   & 22.3   \nl
       &       &         &        &        &$\pm$0.9&    0.6 &        &$\pm$0.5 &$\pm$0.3&$\pm$0.4 &$\pm$0.5\nl
16  02 &\nodata&  3      & \nodata& \nodata&   9.8  &   37.1 & \nodata& 39.8    & 111.2  & 95.6    & 28.1   \nl
       &       &$\pm$1   &        &        &$\pm$0.7&$\pm$0.4&        &$\pm$0.4 &$\pm$0.3&$\pm$0.3 &$\pm$0.4\nl
16  03 &\nodata&  \nodata& \nodata& \nodata&   8.9  &  11.8  & 5.6    &  2.7    &  44.4  & 23.4    & 14.5   \nl
       &       &         &        &        &$\pm$0.3&$\pm$0.2&$\pm$0.3&$\pm$0.5 &$\pm$0.1&$\pm$0.2 &$\pm$0.2\nl
16  04 &1.14   &  4.15   & \nodata& \nodata&  18.60 &   8.33 & 4.30   &  1.2    &  41.48 & 16.42   & 10.60 \nl
16  05 &\nodata&  \nodata& \nodata& \nodata&   3.6  &   8.1  & 3.6    &  7.3    &  59.9  & 22.2    & 14.5  \nl
       &       &         &        &        &$\pm$0.4&   0.2  & 0.4    &  0.3    &   0.1  &  0.1    &  0.2  \nl
16  06 &\nodata&  \nodata& \nodata& \nodata&  14    &  56    & \nodata&  8      &  16    & 46      & 39    \nl
       &       &         &        &        &$\pm$2  &$\pm$1  &        &  $\pm$3 &$\pm$2  &  $\pm$1 &$\pm$1 \nl
22  01 &\nodata&  \nodata& \nodata& \nodata&  26    &  80    & \nodata& 11      & 107    & 21      & 31    \nl
       &       &         &        &        &$\pm$3  &$\pm$2  &        &  $\pm$4 &$\pm$2  &  $\pm$3 &$\pm$2 \nl
23  03 &\nodata&  \nodata& \nodata& \nodata&  26.52 &  52.61 & \nodata& 32.26   & 163.00 &  \nodata&\nodata\nl
23  04 &2.107  &  5.536  & \nodata& \nodata&  22.20&  12.39& 4.74  &  1.66  &  40.74&  \nodata&\nodata\nl
23  05 &\nodata&  \nodata& \nodata& \nodata&$\pm$8.34&  24.58& \nodata&  1.0    &  81.70 &  \nodata&\nodata\nl
       &       &         &        &        &$\pm$0.05&$\pm$0.03&      & $\pm$0.1&$\pm$0.02&        &       \nl
40  01 &\nodata&  \nodata& \nodata& \nodata&  18    &  18    & \nodata&  \nodata&  87    &  \nodata&\nodata\nl
       &       &         &        &        &$\pm$4  &$\pm$4  &        &         &  $\pm$2&         &       \nl
40  04 &\nodata&  \nodata& \nodata& \nodata&   9    &  12    & \nodata& 14      &  65    &  \nodata&\nodata\nl
       &       &         &        &        &$\pm$5  &$\pm$4  &        &  $\pm$4 &  $\pm$2&         &       \nl
40  05 &\nodata&  \nodata& \nodata& \nodata&  11    &  30    & \nodata&  \nodata&  60    &  \nodata&\nodata\nl
       &       &         &        &        &  $\pm$2& $\pm$2 &        &         &$\pm$1  &         &       \nl
42  01 &\nodata&  \nodata& \nodata& \nodata&\nodata &\nodata & \nodata&  \nodata& 178    &  \nodata&\nodata\nl
62  01 &\nodata&  \nodata& \nodata& \nodata&  16    &  16    & \nodata&  8      &  60.1  &  24     &  24   \nl
       &       &         &        &        &$\pm$1  &$\pm$1  &        &  $\pm$2 &$\pm$0.9&$\pm$1   &$\pm$1 \nl
67  02 &\nodata&  \nodata& \nodata& \nodata&  17    &  12    & \nodata& \nodata &  34    &  \nodata& 9     \nl
       &       &         &        &        &$\pm$1  &$\pm$2  &        &         &$\pm$1  &  $\pm$1 & $\pm$2\nl
67  06 &\nodata&  \nodata& \nodata& \nodata&  33    &  23    & \nodata&  \nodata&  37    &  \nodata&\nodata\nl
       &       &         &        &        &$\pm$1  &$\pm$1  &        &         &$\pm$1  &         &       \nl
67  11 &\nodata&  \nodata& \nodata& \nodata& 182    & 211    & \nodata& 99      & 515    &  \nodata&\nodata\nl
       &       &         &        &        &$\pm$31 & $\pm$32&        & $\pm$29 & $\pm$47&         &       \nl
86  01 &\nodata&  \nodata& \nodata& \nodata&  56    &  23    & \nodata& 23      &  62    &  \nodata&\nodata\nl
       &       &         &        &        &$\pm$4  &$\pm$5  &        &  $\pm$5 &$\pm$4  &         &       \nl
86  03 &\nodata&  \nodata& \nodata& \nodata&  16    & 145    & \nodata&  8      &  82    &  \nodata&\nodata\nl
       &       &         &        &        &$\pm$2  &$\pm$1  &        &  $\pm$3 &$\pm$1  &         &       \nl
86  04 &\nodata&  \nodata& \nodata& \nodata&  15    &  23    & \nodata& 15      &  71.0  &  \nodata&\nodata\nl
       &       &         &        &        &$\pm$1  &$\pm$1  &        &  $\pm$1 &$\pm$0.8&         &       \nl
86  07 &\nodata&  \nodata& \nodata& \nodata&  17    & 248    & \nodata& 43      & 123.9  &  \nodata&\nodata\nl
       &       &         &        &        &$\pm$2  &$\pm$1  &        &  $\pm$1 &$\pm$0.9&         &       \nl
86  08 &\nodata&  \nodata& \nodata& \nodata&  10.2  &   4.6  & \nodata&  3.2    &  33.6  & 22.9    &14.6   \nl
       &       &         &        &        &$\pm$0.3&$\pm$0.5&        & $\pm$0.6&$\pm$0.2& $\pm$0.2&$\pm$0.3\nl
86  09 &\nodata&  \nodata& \nodata& \nodata&  13.2  &  10.5  & \nodata&  \nodata&  22.0  &  \nodata&\nodata\nl
       &       &         &        &        &$\pm$0.8&$\pm$0.9&        &         &$\pm$0.6&         &       \nl
87  01 &\nodata&  \nodata& \nodata& \nodata&  24.20 &  28.96 & \nodata& 31.60   &  82.73 &  \nodata&\nodata\nl
87  03 &\nodata&  \nodata& \nodata& \nodata&  50.78 &  22.76 & \nodata&  \nodata& 128.64 &  \nodata&\nodata\nl
87  04 &\nodata&  \nodata& \nodata& \nodata&  22.09 &   9.28 & \nodata&  \nodata&  43.77 &  \nodata&\nodata\nl
88  01 &\nodata&  \nodata& \nodata& \nodata&  14.96 &  45.13 & \nodata& 10.02   &  89.49 &  \nodata&\nodata\nl
88  02 &\nodata&  \nodata& \nodata& \nodata&  23.59 &  36.02 & \nodata&  7.72   & 191.80 &  \nodata&\nodata\nl
88  07 &\nodata&  \nodata& \nodata& \nodata&  18.38 &  23.28 & \nodata&  \nodata&  32.60 &  \nodata&\nodata\nl
90  01 &\nodata&  \nodata& \nodata& \nodata&  16    & 144    & \nodata&  \nodata& 120    & 36      &38     \nl
       &       &         &        &        &$\pm$3  &$\pm$2  &        &         &$\pm$2  &  $\pm$2 & $\pm$2\nl
90  04 &\nodata&  \nodata& \nodata& \nodata&   8    &   8    & \nodata&  8      &  62.7  & 18      &12     \nl
       &       &         &        &        &$\pm$2  &$\pm$1  &        &  $\pm$2 &$\pm$0.6&  $\pm$1 & $\pm$1\nl
90  09 &\nodata&  \nodata& \nodata& \nodata&  27.4  &  44.1  & \nodata&  \nodata&  24.8  &  \nodata&14.2   \nl
       &       &         &        &        &$\pm$0.7&$\pm$0.6&        &         &$\pm$0.7& $\pm$0.6&$\pm$0.9\nl       
97  05 &\nodata&  \nodata& \nodata& \nodata&  23    &  40    & \nodata&  \nodata&  53    &  \nodata&\nodata\nl
       &       &         &        &        &$\pm$3  &$\pm$2  &        &         &$\pm$2  &         &       \nl
97  06 &\nodata&  \nodata& \nodata& \nodata&  16    &  39    & \nodata&  \nodata&  56    &  \nodata&\nodata\nl
       &       &         &        &        &$\pm$2  &$\pm$2  &        &         &$\pm$2  &         &       \nl

\enddata 
\end{deluxetable}
\clearpage
\begin{deluxetable}{lcccclc}
\footnotesize
\tablecaption{Characteristics of HCG emission--line galaxies\label{tbl-2}}
\tablewidth{0pt}
\tablehead{
\colhead{HCG  \#}  &\colhead{B} &\colhead{cz}            & \colhead{M$_{\rm B}$} & \colhead{T} &
\colhead{Activity} & \colhead{EW}             \\
\colhead{}         &\colhead{}  &\colhead{(km s$^{-1}$)} & \colhead{}            & \colhead{}  &
\colhead{Type}     &\colhead{H$\alpha + $[NII]}\\ 
\colhead{}         &\colhead{}  &\colhead{}              &\colhead{}             &\colhead{}   &
\colhead{}         & \colhead{(\AA)} 
}
\startdata
04  01 &13.71  &8035   &-21.44&     5  &  SBNG            & 106\nl
04  03 &15.97  &8242   &-19.23&    -5  &  SBNG            &  44\nl
04  11 &17.75  &6957   &-17.09&\nodata &  HII             & 261\nl
16  01 &12.88  &4073   &-20.79&      2 &  LNR             &  10\nl
16  02 &13.35  &3864   &-20.21&      2 &  Sy2             &  18\nl
16  03 &13.35  &4001   &-20.29&\nodata &  SBNG            &  69\nl
16  04 &13.61  &3859   &-19.95&    10  &  SBNG            & 154\nl
16  05 &13.66  &3934   &-19.94&\nodata &  LNR             &  74\nl
16  06 &15.56  &3972   &-18.06&\nodata &  HII             &  14\nl
22  01 &12.60  &2681   &-20.17&     -5 &  dSy2            &   2\nl
23  02 &15.03  &9899   &-20.57&\nodata &  Sy2             &  28\nl
23  03 &15.21  &4869   &-18.85&      2 &  dLNR            &   5\nl
23  04 &15.81  &4467   &-18.06&     7  &  SBNG            & 181\nl
23  05 &16.14  &5373   &-18.14&     -2 &  dSy2            &   5\nl
40  01 &14.32  &6634   &-20.41&     -5 &  dLNR            &   4\nl
40  04 &15.20  &6362   &-19.44&      1 &  LNR             &  13\nl
40  05 &16.50  &6633   &-18.23&      5 &  Sy2             &  25\nl
42  01 &12.62  &3712   &-20.85&     -5 &  dLNR            &   4\nl
62  01 &13.30  &4259   &-20.47&     -5 &  dLNR            &   5\nl
67  02 &14.57  &7543   &-20.44&     5  &  SBNG            &  28\nl
67  06 &15.66  &7831   &-19.43&\nodata &  SBNG            &  21\nl
67  11 &16.49  &6614   &-18.24&\nodata &  dLNR            &   3\nl
86  01 &14.23  &6013   &-20.29&     -5 &  dLNR            &   2\nl
86  03 &14.81  &5863   &-19.66&     -5 &  dSy2            &   5\nl
86  04 &15.30  &5298   &-18.95&     -2 &  LNR             &   8\nl
86  07 &15.98  &6476   &-18.70&\nodata &  Sy2             &  13\nl
86  08 &16.20  &5508   &-18.13&\nodata &  SBNG            &  90\nl
86  09 &16.26  &6816   &-18.53&\nodata &  SBNG            &  19\nl
87  01 &14.81  &8436   &-20.45&      4 &  dLNR            &   6\nl
87  03 &15.40  &8738   &-19.93&      0 &  dLNR            &   2\nl
87  04 &15.78  &8963   &-19.61&      4 &  SBNG            &  19\nl
88  01 &13.71  &6007   &-20.81&      3 &  dSy2            &   7\nl
88  02 &13.81  &6124   &-20.75&      3 &  dLNR            &  10\nl
90  01 &12.99  &2603   &-19.71&      1 &  Sy2             &   8\nl
90  04 &13.54  &2659   &-19.21&\nodata &  LNR             &  15\nl
90  09 &16.80  &2760   &-16.03&\nodata &  SBNG            &  43\nl
97  05 &15.29  &6379   &-19.36&\nodata &  LNR             &  11\nl
97  06 &15.45  &6666   &-19.29&      6 &  Sy2             &  17\nl
\enddata 
\end{deluxetable}

\clearpage
\begin{deluxetable}{lcccccccccc}
\footnotesize
\tablecaption{Absorption features in HCG AGNs \label{tbl-3}}
\tablewidth{0pt}
\tablehead{
\colhead{HCG  \#}  & \colhead{CaII K} & \colhead{CaII H} & \colhead{H$\delta$}& \colhead{CN}   & \colhead{G--band} &
\colhead{H$\beta$} & \colhead{Mg}     & \colhead{Na}     & \colhead{$\frac{{\rm I(CaII H)}}{{\rm I(CaII K)}}$}     & 
\colhead{Mg$_2$} \\
\colhead{}         & \colhead{(\AA)}  & \colhead{(\AA)}  & \colhead{(\AA)}    &\colhead{(\AA)} & \colhead{(\AA)}   & 
\colhead{(\AA)}    & \colhead{(\AA)}  & \colhead{(\AA)}  & \colhead{}         & \colhead{}
}
\startdata
16  01&  \nodata& \nodata&    2.6& 1.9& 4.9&     4.1& 3.7&    5.7& \nodata& 0.21\nl
16  02&  \nodata& \nodata&    1.4& 2.9& 8.4& \nodata& 4.9&    7.6& \nodata& 0.33\nl
16  05&  \nodata& \nodata&    6.0& 2.9& 2.7&     5.7& 1.7&    4.4& \nodata&\nodata\nl
40  04&     6.28&    2.62&\nodata& 4.1& 5.1& \nodata& 4.6&    4.8&    1.0& 0.19\nl
40  05&  \nodata& \nodata&    2.6& 4.7& 6.6& \nodata& 4.6&\nodata& \nodata& 0.22\nl
86  04&  \nodata& \nodata&    2.7& 2.4& 8.9&     2.6& 5.5&    5.2& \nodata& 0.33\nl
86  07&     9.08&    5.78&    2.4& 2.0& 4.3&     4.0& 3.5&    3.6&    1.1& 0.13\nl
90  01&  \nodata& \nodata&    2.3& 4.1& 7.1&     4.0& 4.3&    3.7& \nodata& 0.17\nl
90  04&  \nodata& \nodata&    1.7& 4.8& 5.5&     2.9& 4.4&    4.2& \nodata& 0.19\nl
\enddata 
\end{deluxetable}

\clearpage
\begin{deluxetable}{lcccccccccc}
\footnotesize
\tablecaption{Absorption features in HCG LLAGNs \label{tbl-4}}
\tablewidth{0pt}
\tablehead{
\colhead{HCG  \#}  & \colhead{CaII K} & \colhead{CaII H} & \colhead{H$\delta$}& \colhead{CN}   & \colhead{G--band} &
\colhead{H$\beta$} & \colhead{Mg}     & \colhead{Na}     & \colhead{$\frac{{\rm I(CaII H)}}{{\rm I(CaII K)}}$}     & 
\colhead{Mg$_2$} \\
\colhead{}         & \colhead{(\AA)}  & \colhead{(\AA)}  & \colhead{(\AA)}    &\colhead{(\AA)} & \colhead{(\AA)}   & 
\colhead{(\AA)}    & \colhead{(\AA)}  & \colhead{(\AA)}  & \colhead{}         & \colhead{}
}
\startdata
22  01&\nodata&\nodata& 2.2& 6.0& 7.0&  2.9& 4.9& 5.4&\nodata& 0.33\nl
23  03&\nodata&\nodata& 2.2& 4.1& 6.4&  2.2& 4.3& 4.6&\nodata& 0.29\nl
23  05&\nodata&\nodata& 3.8& 2.4& 2.3&  5.6& 3.0& 4.2&\nodata& 0.13\nl
40  01&  12.0&   8.6&   2.9& 1.8& 6.6&  3.6& 4.3& 6.1&   1.1& 0.26\nl
42  01&\nodata&\nodata& 1.8& 2.5& 7.5&  2.1& 4.8& 9.3&\nodata& 0.34\nl
62  01&\nodata&\nodata& 1.8& 2.8& 6.2&  1.7& 4.8& 6.9&\nodata& 0.33\nl
67  11&   9.3&   7.8&   2.4& 2.5& 6.6&  2.0& 4.8& 5.2&   1.0& 0.34\nl
86  01&  11.2&   9.2&   2.0& 2.5& 6.1&  2.5& 4.9& 8.9&   1.1& 0.35\nl
86  03&   9.9&   6.6&   1.6& 2.6& 7.0&  2.8& 5.1& 5.8&   1.0& 0.31\nl
87  01&   5.7&   6.3&   1.5& 3.8& 4.9&  1.3& 3.8& 6.1&   1.0& 0.25\nl
87  03&  12.2&   8.2&   2.1& 2.9& 7.2&  1.9& 5.4& 5.8&   1.1& 0.31\nl
88  01&  13.0&   9.0&   1.9& 4.4& 5.6&  5.1& 3.9& 5.4&   1.1& 0.25\nl
88  02&  12.3&  11.1&   2.0& 2.4& 7.5&  3.0& 4.4& 4.0&   1.1& 0.23\nl
\enddata 
\end{deluxetable}

\clearpage
\begin{deluxetable}{lcccccccccc}
\footnotesize
\tablecaption{Absorption features in HCG starbursts \label{tbl-5}}
\tablewidth{0pt}
\tablehead{
\colhead{HCG  \#}  & \colhead{CaII K} & \colhead{CaII H} & \colhead{H$\delta$}& \colhead{CN}   & \colhead{G--band} &
\colhead{H$\beta$} & \colhead{Mg}     & \colhead{Na}     & \colhead{$\frac{{\rm I(CaII H)}}{{\rm I(CaII K)}}$}     & 
\colhead{Mg$_2$} \\
\colhead{}         & \colhead{(\AA)}  & \colhead{(\AA)}  & \colhead{(\AA)}    &\colhead{(\AA)} & \colhead{(\AA)}   & 
\colhead{(\AA)}    & \colhead{(\AA)}  & \colhead{(\AA)}  & \colhead{}         & \colhead{}
}
\startdata
04  01&    2.5&    2.2&     em&   1.6&   2.9&    em&   1.6&   1.4&   1.0&\nodata\nl
04  03&    3.8&    5.9&    4.4&   2.0&   2.7&   4.1&   2.8&   1.4&   1.0&   0.09\nl
04  11&\nodata&\nodata&     em&   3.7&\nodata&   em&\nodata&\nodata&\nodata&   0.27\nl
16  03&\nodata&\nodata&    5.8&   2.5&   3.0&   5.2&   2.4&   2.4&\nodata&\nodata\nl
16  04&\nodata&\nodata&    3.3&   1.0&   0.6&    em&   1.4&   2.9&\nodata&\nodata\nl
16  06&\nodata&\nodata&    3.2&   3.5&   4.7&   3.1&\nodata&\nodata&\nodata&\nodata\nl
23  04&\nodata&\nodata&     em&   1.8&   1.5&    em&   1.7&   0.8&\nodata&\nodata\nl
63  06&\nodata&\nodata&     em&   2.3&   4.3&    em&   2.6&   4.4&\nodata&   0.09\nl
67  02&\nodata&\nodata&\nodata&   4.7&   5.6&    em&   4.9&   2.3&\nodata&   0.21\nl
86  05&   10.7&    5.8&    2.8&\nodata&  5.2&   3.7&   3.6&   4.5&   0.9&   0.21\nl
86  08&\nodata&\nodata&    4.2&   1.5&\nodata&   em&\nodata&   3.8&\nodata&\nodata\nl
86  09&\nodata&\nodata&    1.9&   5.7&   5.1&   1.6&   4.7&   3.9&\nodata&   0.19\nl
87  02&\nodata&\nodata&\nodata&   4.0&\nodata&   em&   2.5&\nodata&\nodata&\nodata\nl
87  05&\nodata&\nodata&\nodata&   2.2&\nodata&   em&\nodata&\nodata&\nodata&\nodata\nl
87  07&    3.8&    4.2&    4.4&   2.0&   2.2&   4.3&   2.1&   2.3&   1.0&\nodata\nl
88  07&    6.2&    7.8&     em&   4.8&   7.0&   4.4&   3.5&   3.2&   1.3&   0.34\nl
\enddata
\end{deluxetable}

\clearpage
\begin{deluxetable}{lccccccccccc}
\footnotesize
\tablecaption{Absorption features in HCG galaxies without emission \label{tbl-6}}
\tablewidth{0pt}
\tablehead{
\colhead{HCG  \#}  & \colhead{CaII K} & \colhead{CaII H} & \colhead{H$\delta$}& \colhead{CN}   & \colhead{G--band} &
\colhead{H$\beta$} & \colhead{Mg}     & \colhead{Na}     & \colhead{$\frac{{\rm I(CaII H)}}{{\rm I(CaII K)}}$}     & 
\colhead{Mg$_2$}   & \colhead{T} \\
\colhead{}         & \colhead{(\AA)}  & \colhead{(\AA)}  & \colhead{(\AA)}    &\colhead{(\AA)} & \colhead{(\AA)}   & 
\colhead{(\AA)}    & \colhead{(\AA)}  & \colhead{(\AA)}  & \colhead{}         & \colhead{}     & \colhead{}
}
\startdata
19  01&\nodata&\nodata& 2.3& 2.5& 6.5&  3.1& 4.5& 3.6&\nodata& 0.27&   -5\nl
22  03&\nodata&\nodata& 3.7& 2.7& 4.1&  4.9& 3.6& 1.9&\nodata& 0.16&\nodata\nl
22  04&\nodata&\nodata& 3.8& 2.1& 6.6&  4.1& 3.6& 2.9&\nodata& 0.17&    1\nl
40  02&  12.6&   7.7&   2.7& 2.7& 5.9&  3.5& 4.1& 6.0&    1.0& 0.26&   -2\nl
42  02&\nodata&\nodata&\nodata& 3.5& 6.9&2.4& 4.7&5.4&\nodata& 0.30&   -2\nl
42  04&\nodata&\nodata& 2.8& 3.6& 6.4&  5.1& 4.0& 5.7&\nodata& 0.29&   -5\nl
48  01&\nodata&\nodata& 2.7& 3.4& 8.5&  3.0& 4.3& 5.0&\nodata& 0.32&   -5\nl
62  02&\nodata&\nodata& 2.4& 1.9& 6.2&  3.3& 4.9& 6.5&\nodata& 0.32&   -2\nl
62  03&\nodata&\nodata& 1.4& 2.7& 6.0&  1.9& 4.9& 4.1&\nodata& 0.27&   -2\nl
62  04&\nodata&\nodata& 2.2& 3.3& 7.2&  2.5& 4.5& 3.9&\nodata& 0.29&     \nl
62  08&\nodata&\nodata& 2.8& 5.2& 8.0&  2.8& 4.7& 3.5&\nodata& 0.26&\nodata\nl
63  09&   4.4&   3.6&   2.0& 2.5& 5.5&  1.7& 4.9& 5.3&    1.1& 0.27&\nodata\nl
67  01&  10.5&   6.2&   1.4& 3.0& 7.1&  1.7& 4.4& 5.5&    1.1& 0.30&   -5\nl
67  03&  10.6&   6.9&   2.5& 3.5& 7.0&  2.5& 5.1& 4.3&    1.0& 0.28&\nodata\nl
86  02&  12.1&   9.1&   2.6& 1.4& 7.2&  3.5& 4.8& 7.5&    1.1& 0.33&\nodata\nl
86  06&  11.9&   8.2&   1.7& 2.9& 6.7&  2.4& 4.0& 4.9&    1.0& 0.26&   -2\nl
90  03&\nodata&\nodata& 1.7& 2.6& 8.3&  1.5& 5.0& 4.9&\nodata& 0.29&   -5\nl
97  02&  10.8&   8.2&   2.8& 2.3& 5.8&  4.9& 3.7& 5.3&    1.1& 0.24&\nodata\nl
97  10&  14.2&   9.8&   2.9& 3.3& 7.5&  3.1& 4.6& 5.2&    1.0& 0.31&\nodata\nl
97  11&   6.6&   6.5&   2.8& 3.4& 6.6&  3.0& 3.7& 4.0&    1.1& 0.25&\nodata\nl
\enddata
\end{deluxetable}

\clearpage
\begin{deluxetable}{lccc}
\footnotesize
\tablecaption{Electron densities and metallicities of HCG Starbursts \label{tbl-7}}
\tablewidth{0pt}
\tablehead{
\colhead{HCG  \#}                     & \colhead{$\frac{\lambda 6717}{\lambda 6731}$} &\colhead{N$_e$}       & 
\colhead{12 $+$ Log(O/H)} \\
\colhead{}                            &\colhead{}                                     &\colhead{(cm$^{-3}$)} &
\colhead{}                             
}
\startdata
04  01& 1.5    & 110   & 8.9\nl
04  03& 1.4    &  70   & 9.0\nl 
04  11& 1.0    & 800   & 8.1\nl
16  03& 1.6    & $<$10 & 8.6\nl 
16  04& 1.6    & $<$10 & 9.0\nl
16  06& 1.2    & 300   & 8.3\nl
19  05& 1.3    & 100   & 8.8\nl
22  07&\nodata &\nodata& 9.1\nl
23  04&\nodata &       & 8.9\nl
23  06& 1.5    & $<$10 & 8.8\nl
48  19& 1.4    &  70   & 8.9\nl
48  25& 1.6    & $<$10 & 8.5\nl
63  06&\nodata &\nodata& 9.0\nl
64  22&\nodata &\nodata& 8.8\nl
67  02&\nodata &\nodata& 8.8\nl
67  06&\nodata &\nodata& 8.8\nl
86  08& 1.6    & $<$10 & 9.0\nl 
86  09&\nodata &\nodata& 8.7\nl
87  02& 1.1    & 600   & 9.0\nl
87  05& 1.2    & 300   & 8.5\nl
87  07&\nodata &\nodata& 8.7\nl
88  07&\nodata &\nodata& 8.6\nl
90  09&\nodata &\nodata& 8.6\nl
\enddata 
\end{deluxetable}

\clearpage

\begin{references} 
\reference{} Abraham, R. G., Smecker--Hane, T. A., Hutchings, J. B., Carlberg, R. G., 
Yee, H. K. C., Ellingson, E., Morris, S., Oke, J. B., Rogler, M. 1997, \apj, 476, 7
\reference{} Bahcall, J. N., Kirhakos, S., Saxe, D. H., Schneider, D. P. 1997, \apj, 479, 642 
%\reference{} Baldwin, J. A., Phillips, M. M., Terlevich, R. 1981, \pasp, 93, 5 
\reference{} Barbaro, G., Poggianti, B. M., 1997, \aap, 324, 490 
\reference{} Baugh, C. M., Cole, S., Frenk, C. S., 1996, \mnras, 282, 27 
\reference{} Bender, R., Burstein, Faber, S. M. 1993, \apj, 411, 137 
\reference{} Bender, R., Saglia, R. P., Ziegler, B. 1997, in The Early Universe with the
VLT, eds.  J.  Bergeron et al., Springler, in press 
\reference{} Brodie, J. P., Huchra, J. P. 1990, \apj, 362, 503
\reference{} Butcher, H., Oemler, A. 1978, \apj, 219, 18
\reference{} Caldwell, N., Rose, J. A., Franx, M., Leonardi, A. J. 1996, \aj, 111, 78 
\reference{} Caldwell, N., Rose, J.  1997, \aj, 113, 492 
\reference{} Carlberg, R.  G.  1990, \apj, 350, 505
%\reference{} Charlot, S., Bruzual, G. A. 1991, \apj, 367, 126
\reference{} Cid Fernandes, R., Storchi Bergmann, T. S., Schmitt, H. R. 1998, \mnras, in press (astro-ph/9801309)
\reference{} Clements, D. L., Couch, W. J. 1996, \mnras, 280, 43
\reference{} Cowie, L. L., Songaila, A., Hu, E. M., Cohen, J. G. 1996, \aj, 112, 839 
\reference{} Coziol, R.  1996, \aap, 309, 345
\reference{} Coziol, R.  1997, in Young Galaxies and QSO Absorption--line Systems, 
eds. S. M. Viegas, R.  Gruewald, R. R. R. de Carvalho, PASP conf. series, Vol 114, p. 63
\reference{} Coziol, R., Demers, S., Pe\~na, M., Barneoud, R. 1994, \aj, 108, 405 
\reference{} Coziol, R., Contini, T., Davoust, E., Consid\`ere, S. 1997a, \apj, 481, L67 
\reference{} Coziol, R., Ribeiro, A.  L.  B., de Carvalho, R.  R., 
Capelato, H. V.  1997b, \apj, 493, 563, Paper~I
\reference{} Coziol, R., Demers, S., Barneoud, R., Pe\~na, M.  1997c, \aj, 113, 1548 
%\reference{} de Carvalho, R. R., Ribeiro, A. L. B., Zepf, S. E. 1994, \apjs, 93, 47
\reference{} de Carvalho, R. R., Ribeiro, A. L. B., Capelato, H. V., 
Zepf, S. E. 1997, \apjs, 110, 1 
\reference{} Diafero, A., Geller, M. J., Ramella, M. 1994, \aj, 107, 868
\reference{} Diafero, A., Geller, M. J., Ramella, M. 1995, \aj, 109, 2293
\reference{} Dressler, A.  1980, \apj, 236, 351 
%\reference{} Dressler, A., Shectman, S.  A.  1987, \aj, 94, 899 
\reference{} Dressler, A., Smail, I. 1997, in HST and the High Redshift Universe, 
eds. N. R. Tanvir, A. Aragon--Salamanca, J. V. Wall, World Scientific Press, in press 
\reference{} Ellis, R.  S., Smail, I., Dressler, A., Couch, W. J., Oemler, Jr. A., 
Butcher, H., Sharples, R.  M.  1997, \apj, 483, 582 
\reference{} Faber, S.  M., Wegner, G., Burstein, D., Davies, R.  L., Dressler, A., 
Lynden--Bell, D., Terlevich, R.  J.  1989, \apj, 71, 173 
\reference{} Gonz\'alez Delgado, R. M., P\'erez, E., Tadhunter, C., Vilchez, J. M., 
Rodr\'{\i}guez--Espinoza, J. M. 1997, \apjs, 108, 155
\reference{} Governato, F., Tozzi, P., Cavalieri, A. 1996; \apj, 458, 18
\reference{} Hamilton, D., 1985, \apj, 297, 371 
\reference{} Hickson, P.  1982, \apj, 255, 382
\reference{} Huchtmeier, W. K., 1997, \aap, 325, 473
\reference{} Jablonka, P., Alloin, D.  1995, \aap, 298, 361
\reference{} Jablonka, P., Martin, P., Arimoto, N.  1996, \aj, 112, 1415
\reference{} Jorgensen, I. 1997, \mnras, 288, 161 
\reference{} Kauffmann, G., Charlot, S., White, S.  D.  M. 1996, \mnras, 283, 117 
\reference{} Kennicutt, Jr.  R.  C., Kent, S. M. 1983, \aj, 88, 1094 
\reference{} Kennicutt, Jr.  R.  C.  1992, \apjs, 79, 255 
\reference{} Kennicutt, Jr.  R.  C.  1994, in The Evolution of Galaxies and their 
Environment, eds.  J.  M.  Shull, H. A.  Thronson, Kluwer 
\reference{} Larson, R. B. 1990, \pasp, 102, 709 
\reference{} Larson, R. B., Tinsley, B. M., Caldwell, C. N. 1980, \apj, 237, 692 
\reference{} Lavery, R. J., Seitzer, P., Suntzeff, N. B., Walker, A. R., 
Da Costa, G.  S.  1996, \apj, 467, L1 
\reference{} Leon, S., Combes, F., Menon, T. K. 1998, \aap, 330, 37
\reference{} Leonardi, A. J., Rose, J. A. 1996, \aj, 111, 182
\reference{} Lowenthal, J. D., Koo, D. C., Guzman, R., Gallego, J., Phillips, A. C., 
Faber, S. M., Vogt, N. P., Illingworth, G. D., Gronwall, C. 1997, \apj, 481, 673
\reference{} Malkan, M. A., Gorjian, V., Tam, R. 1998, \apjs, 117, in press (astro-ph/9803123)
\reference{} Mamon, G. A. 1994, in Clusters of Galaxies, eds.  F.  Durret, A.  Mazure, 
Tran Than Van, J., \'editions fronti\`eres, p.  297 
\reference{} Mendes de Oliveira, C., Hickson, P. 1994, \apj, 427, 684 
\reference{} Neuschaefer, L. W., IM, M., Ratnatunga, K. U., Griffiths, R. E., 
Casertano, S. 1997, \apj, 480, 59 
\reference{} Omont, A., Petitjean, P., Guilloteau, S., McMahon, S., Solomon, R. G.,
Pecontal, E. 1996, Nature, 382, 428 
\reference{} OOsterloo, T., Iovino, A. 1997, PASA, 14, 48
\reference{} Osterbrock, D. E. 1989, in Astrophysics of Gaseous Nebulae and Active Galactic Nuclei,
University Science Book, Mill Valley, California 
\reference{} Osterbrock, D. E., Mathews, G. 1986, \araa, 24, 171
\reference{} Ostriker, J. P., Lubin, L. M., Hernquist, L. 1995, \apj, 444, L61
%\reference{} Pagel, B. E. J., Edmunds, M. G., Blackwell, D. E., Chun, M.  S., 
%Smith, G. 1979, \mnras, 189, 95 
\reference{} Poggianti, B. M., Barbaro, G. 1996, \aap, 314, 379 
\reference{} Ponman, T. J., Borner, P. D. J., Ebeling, H., B\"{o}ringer, H.,
1996, \mnras, 283, 690 
\reference{} Rakos, K. D., Maindl, T. I., Schombert, J. M. 1996, \apj, 466, 122 
\reference{} Rauch, M., Haehnelt, M. G., Steinmetz, M. 1997, \apj, 481, 601 
%\reference{} Ribeiro, A. L. B., de Carvalho, R. R., Coziol, R., Capelato, H.
%V., Zepf, S.  E.  1996, \apj, 463, L5 
\reference{} Ribeiro, A. L. B., de Carvalho, R. R., Capelato, H. V., Zepf, S. E.  
1998, \apj, 497, 72 
\reference{} Rose, J.  A.  1985, \aj, 90, 1927
\reference{} Rubin, V.  C., Hunter, D.  A., Kent Ford, Jr.  W.  1991, \apjs, 76, 153 
\reference{} Sanders, D.  B., Soifer, B.  T., Elias, J. H., Madore, B. F., 
Matthews, K., Neugebaueur, G., Scoville, N. Z. 1988, \apj, 325, 74 
\reference{} Schade, D., Barrientos, L. F., Lopez--Cruz, O. 1997, \apj, 477, 17 
\reference{} Simard, L., Pritchet, C. J. 1996, \apj, in press (astro--ph/9606006) 
\reference{} Skillman, E. D., Kennicutt, Jr. R. C., Shields, G. A., Zaritsky, D.  
1996, \apj, 462, 147 
\reference{} Stein, P.  1997, \aap, 480, 395
%\reference{} Steinmetz, M. 1997, in The Early Universe with the VLT,
%eds.  J.  Bergeron te al., Springler, in press 
\reference{} Sofue, Y., Habe, A. 1992, \pasj, 44, 325 
\reference{} Storchi--Bergmann, T., Calzetti, D., Kinney, A. L. 1994, \apjs, 429, 572 
\reference{} Storchi--Bergmann, T., Kinney, A. L., Challis, P. 1995, \apjs, 98, 103  
\reference{} Tinsley, B.  M., Larson, R.  B.  1979, \mnras, 186, 503 
\reference{} Toomre, A., Toomre, J. 1972, \apj, 178, 623
%\reference{} Veilleux, S., Osterbrock, D.  E.  1987, \apjs, 63, 295
\reference{} Veilleux, S., Kim, D.  -C., Sanders, D.  B., Mazzarella,
J.  M., Soifer, B.  T.  1995, \apjs, 98, 171
\reference{} Verdes-Montenegro, L., Yun, M. S., Perea, J., del Olmo, A.,
Ho, P. T. P. 1998, \apj, in press (astro-ph/9709121) 
\reference{} Weedman, D. W. 1983, \apj, 266, 479 
%\reference{} Whitmore, B. C. 1990, in Cluster of Galaxies, eds. W. R. Oegerle, 
%M. J. Fitchett, L. Danly, Cambridge University Press, p.  139 
\reference{} Worthey, G., Faber, S. M., Gonz\'alez, J. J. 1992, \apj, 398, 69 
\reference{} Worthey, G., Faber, S. M., Gonzalez, J. J., Burstein, D. 1994, \apjs, 94, 687 
\reference{} Zabludoff, A. I., Zaritsky, D., Lin, H., Tucker, D., Hashimoto, Y., 
Shectman, S. A., Oemler, A., Kirshner, R. P. 1996, \apj, 466, 104 
\reference{} Zaritsky, D., Zabludoff, A. I., Willick, J.  A.  1996, \aj, 110, 160 
\reference{} Zepf, S. E., Whitmore, B.  C.  1991, \apj, 383, 542 
\reference{} Zepf, S. E. 1993, \apj, 407, 448
\reference{} Zepf, S. E., de Carvalho, R. R., Ribeiro, A. L. B. 1997, \apj, 488, L11 

\end{references}
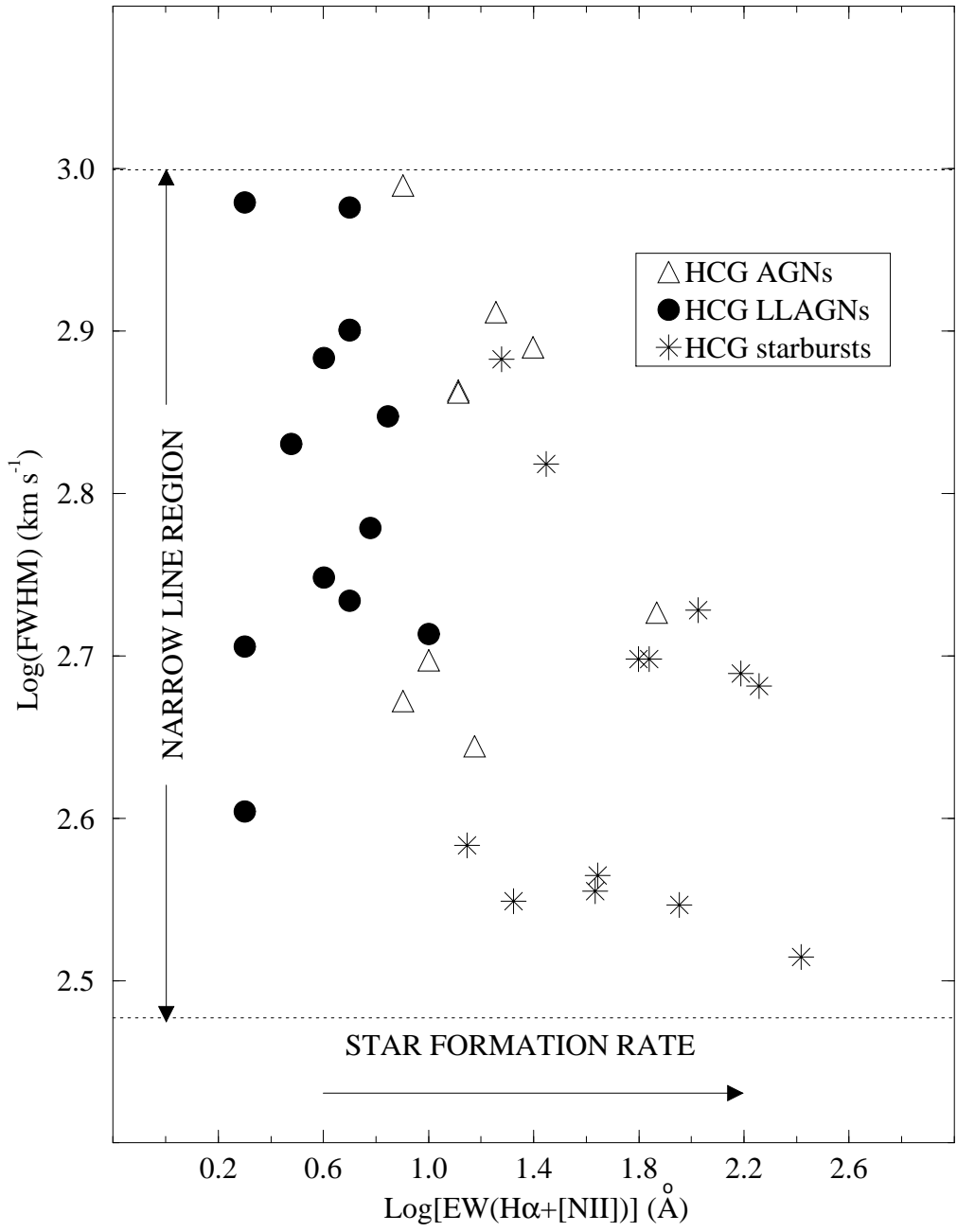
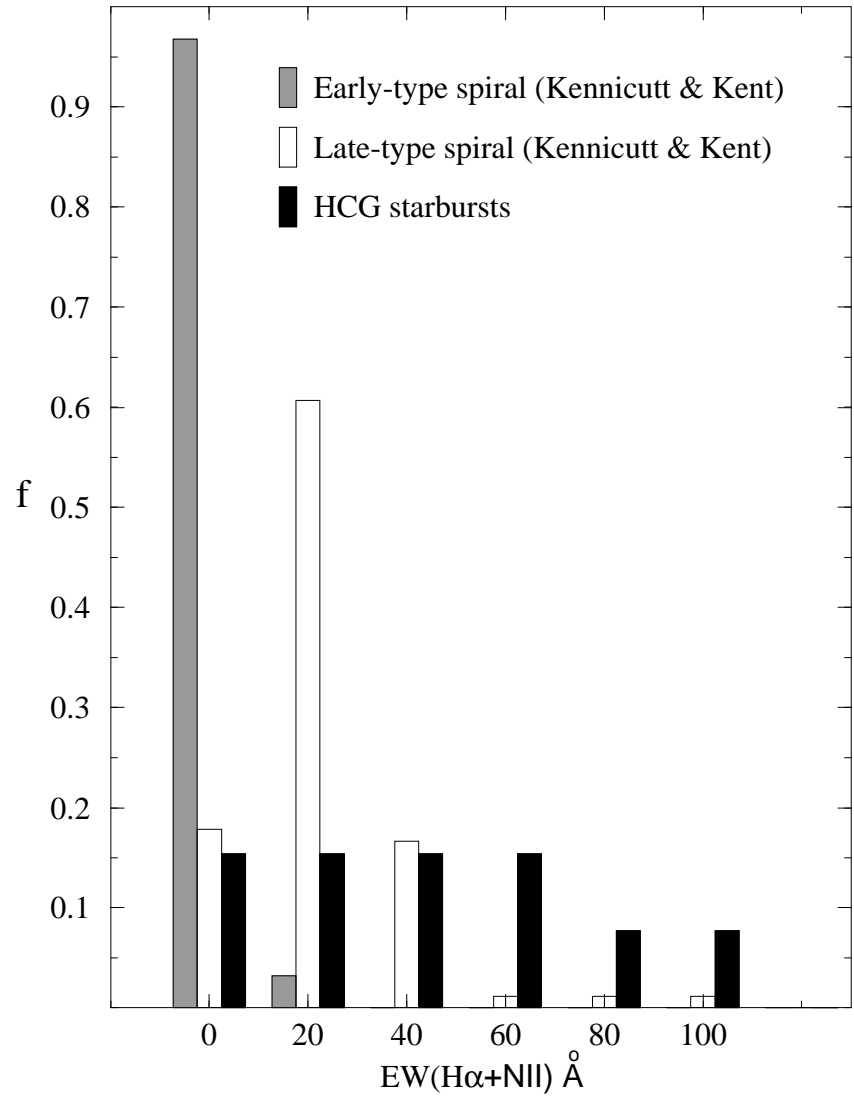
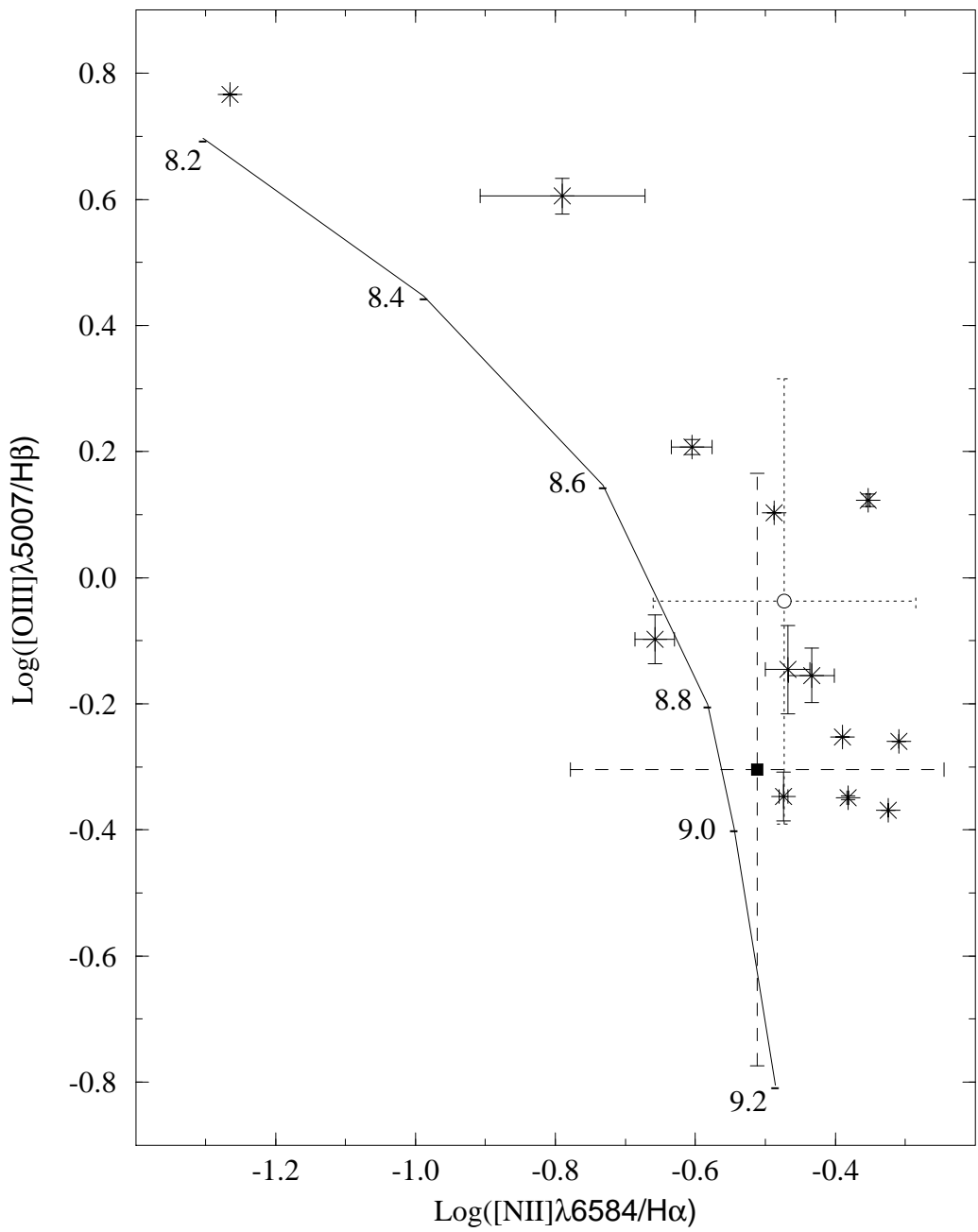
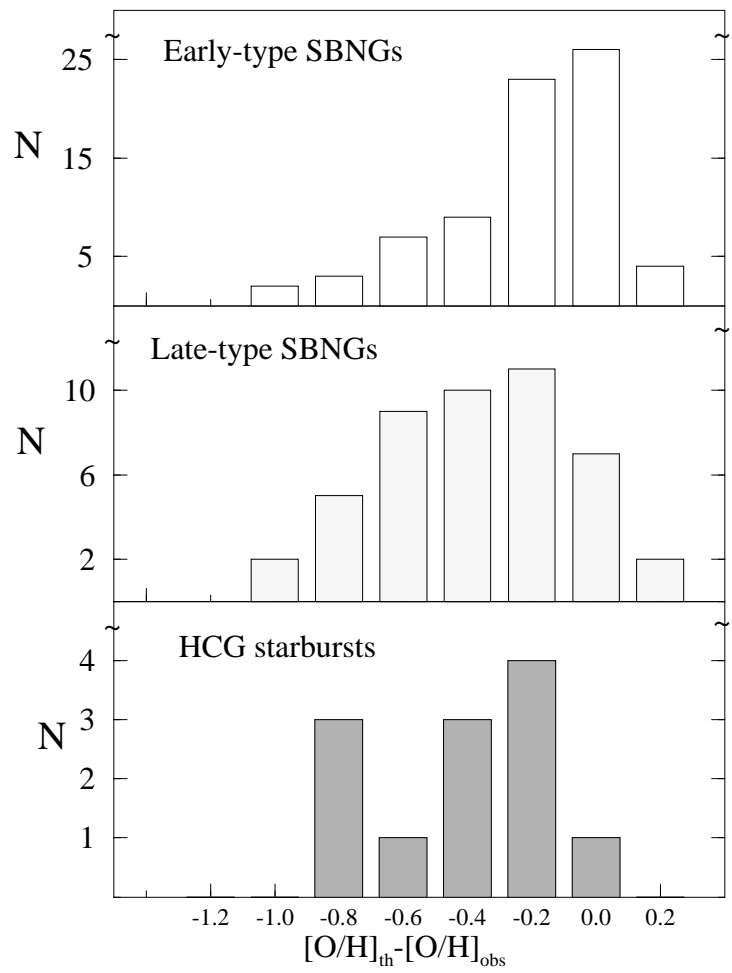
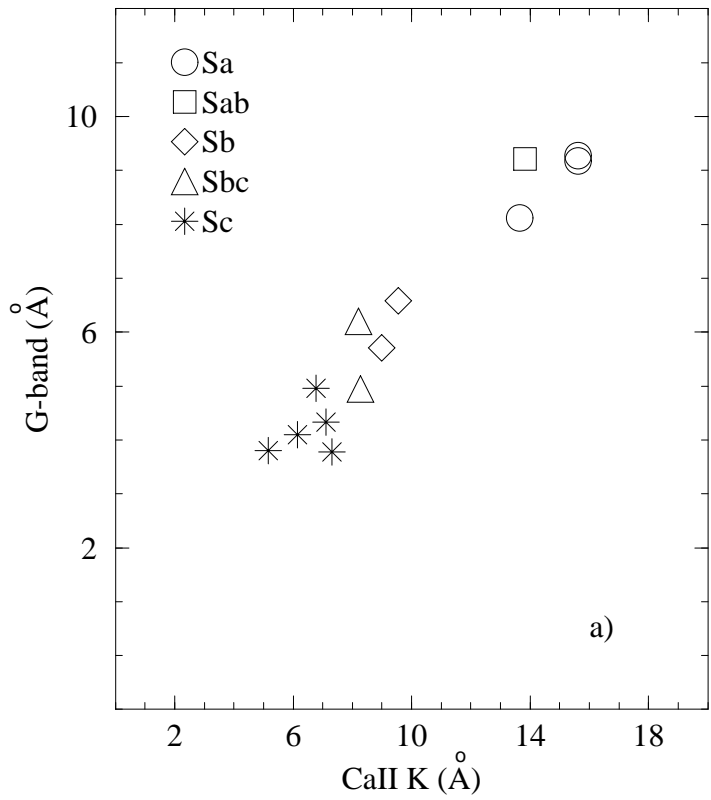
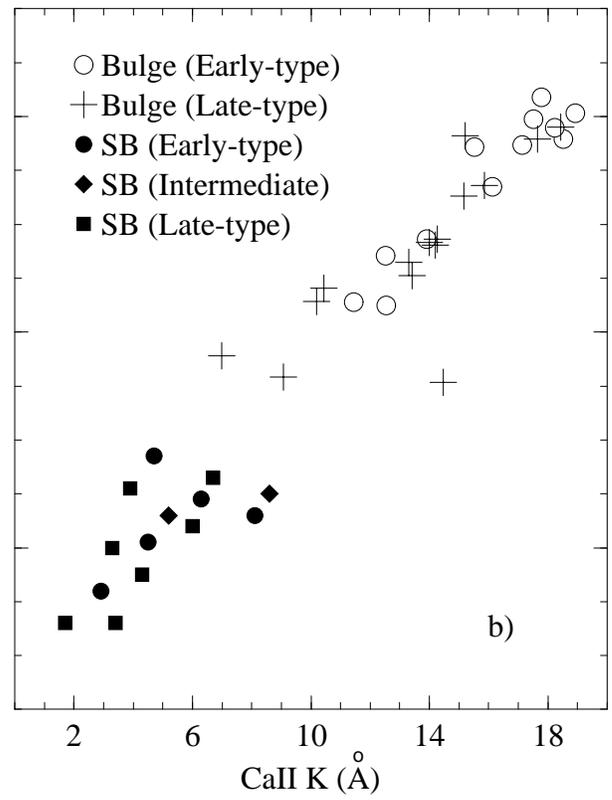
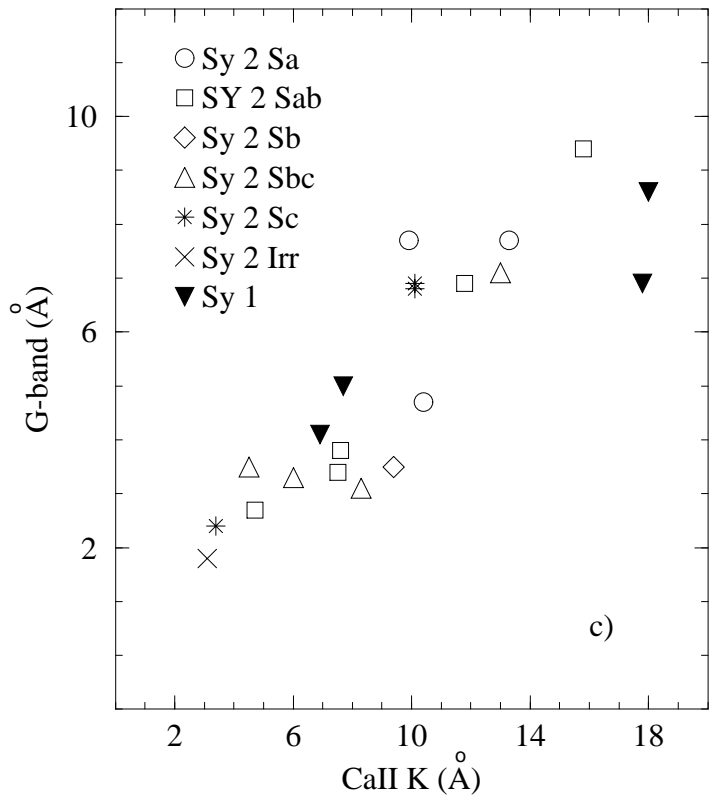
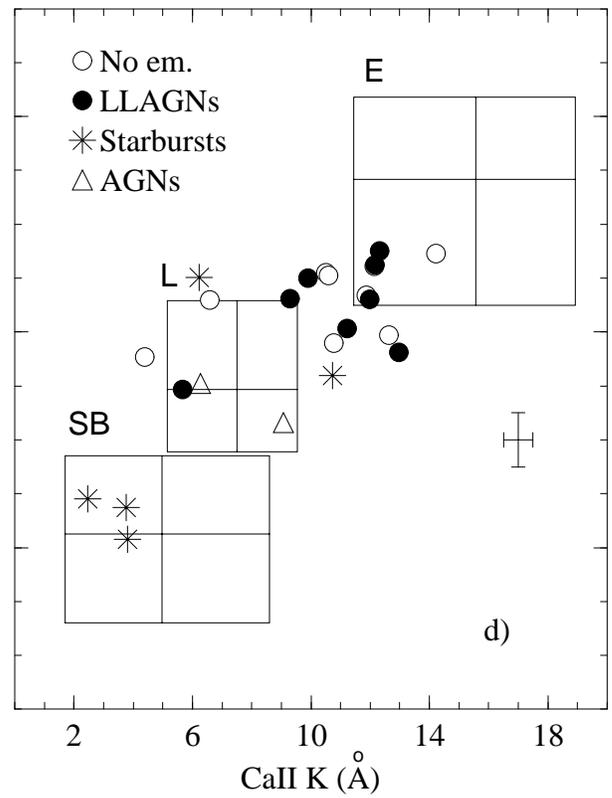
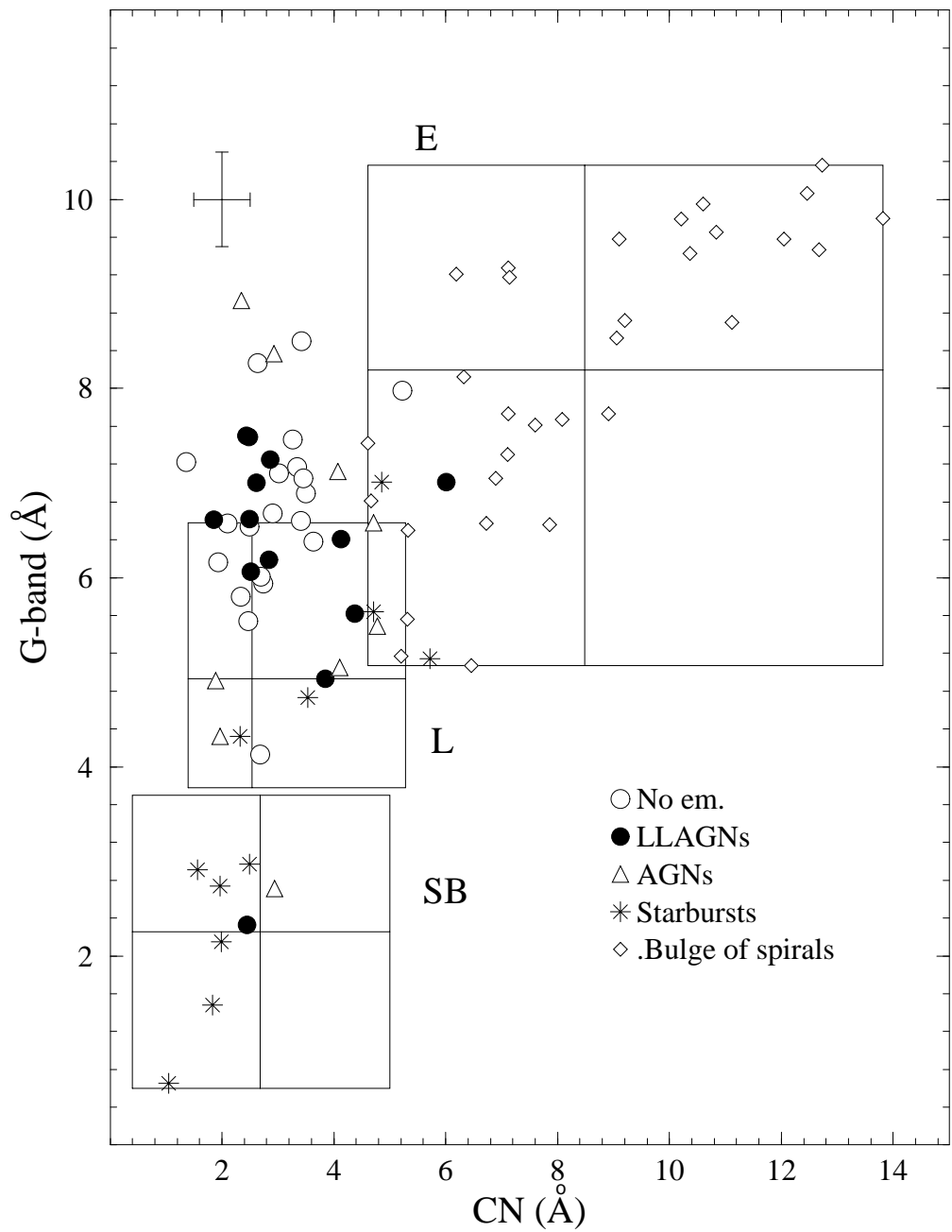
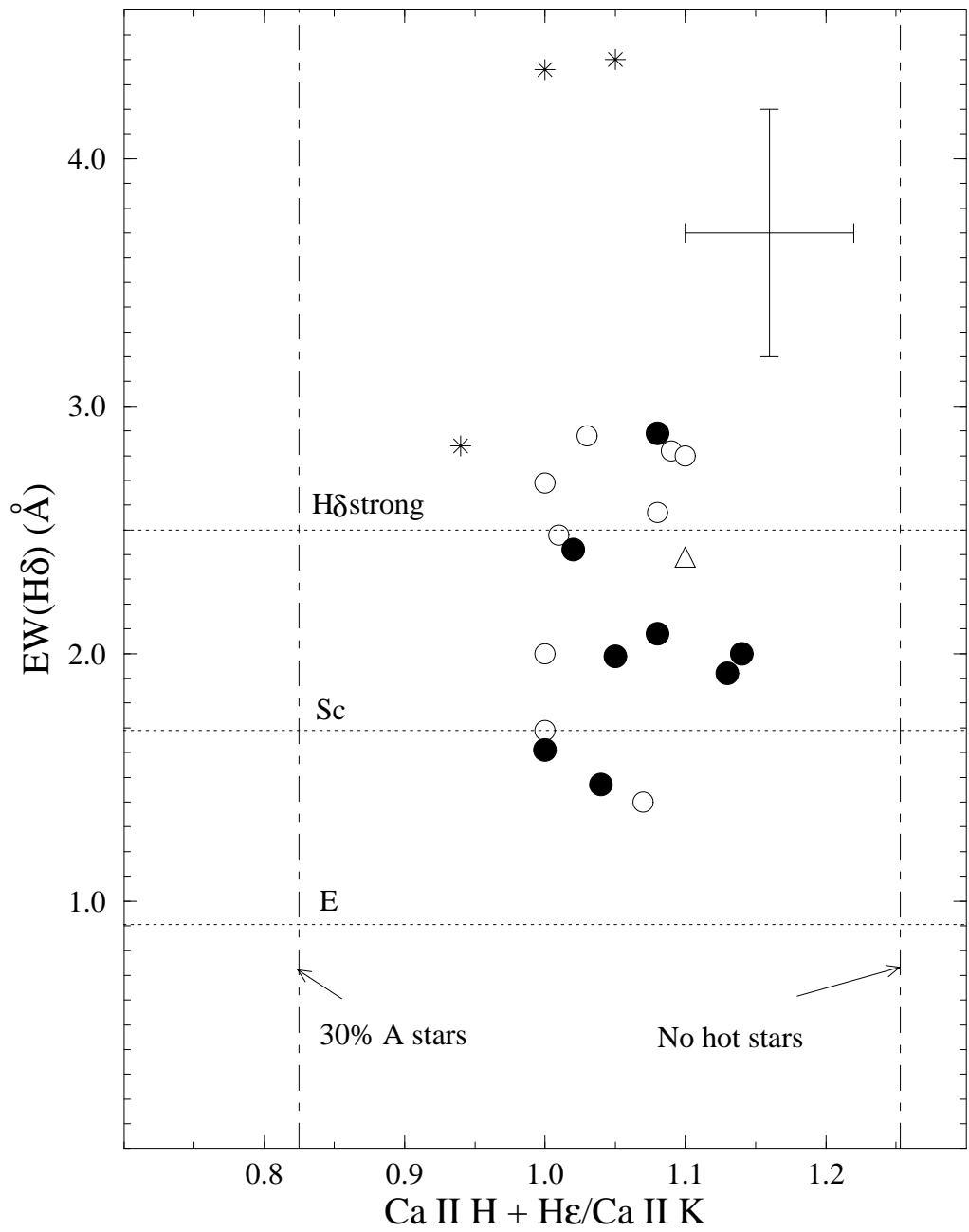
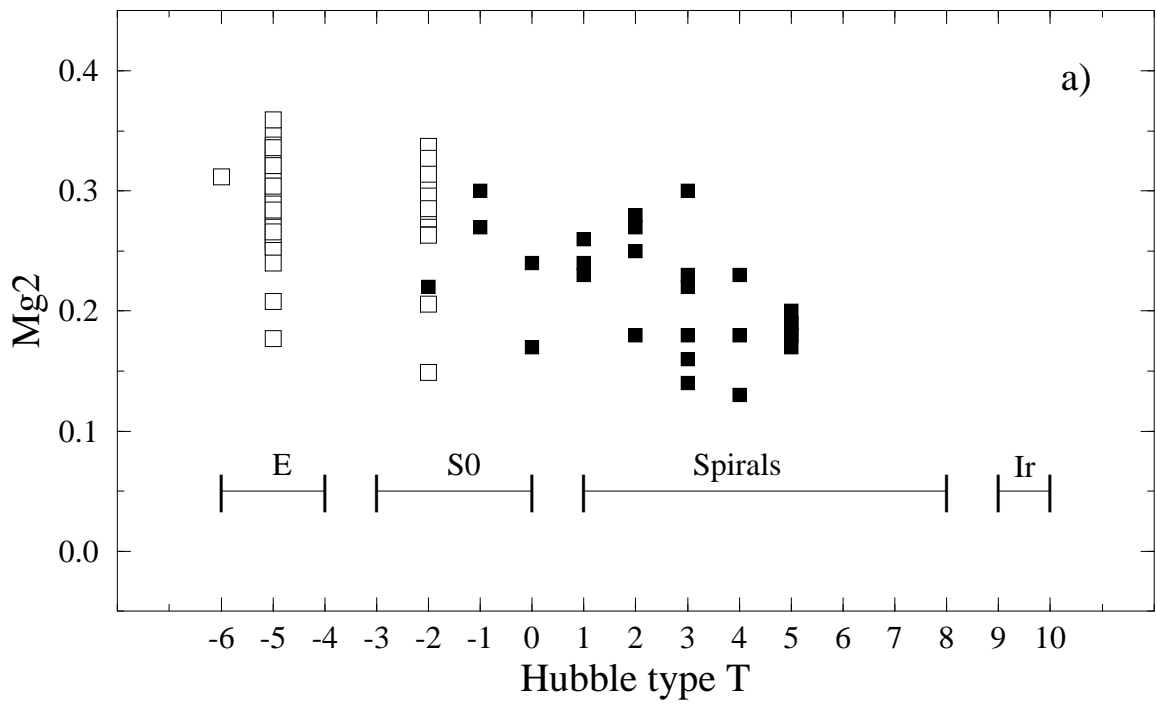
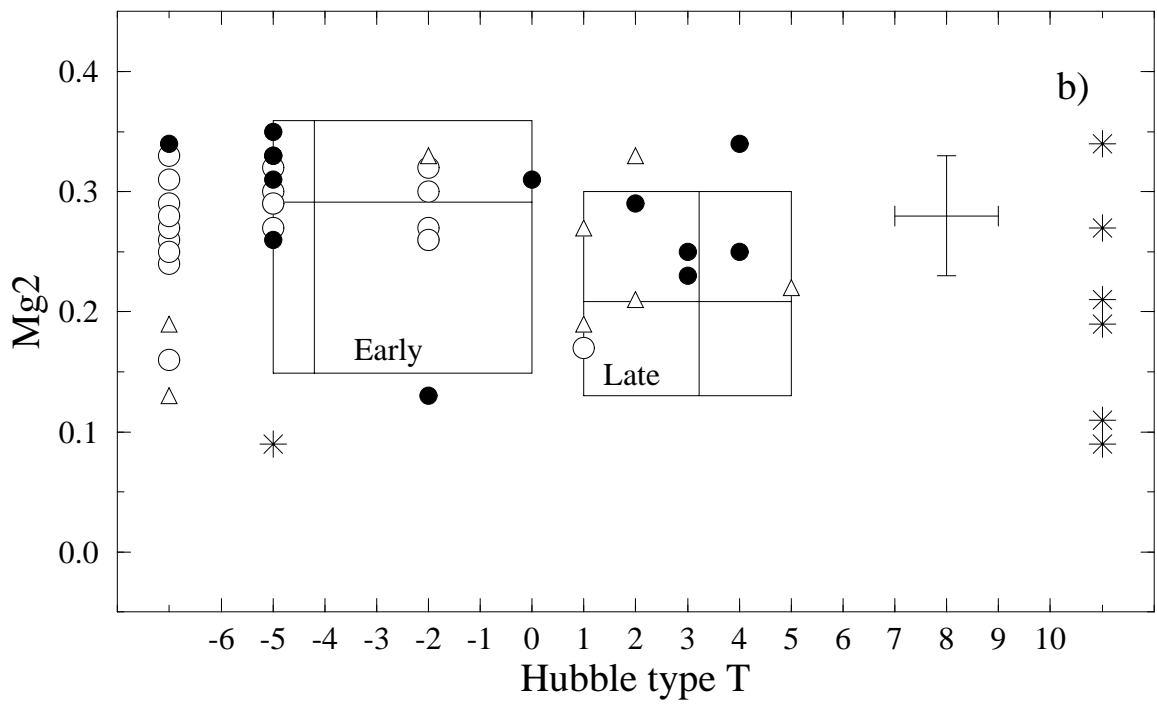
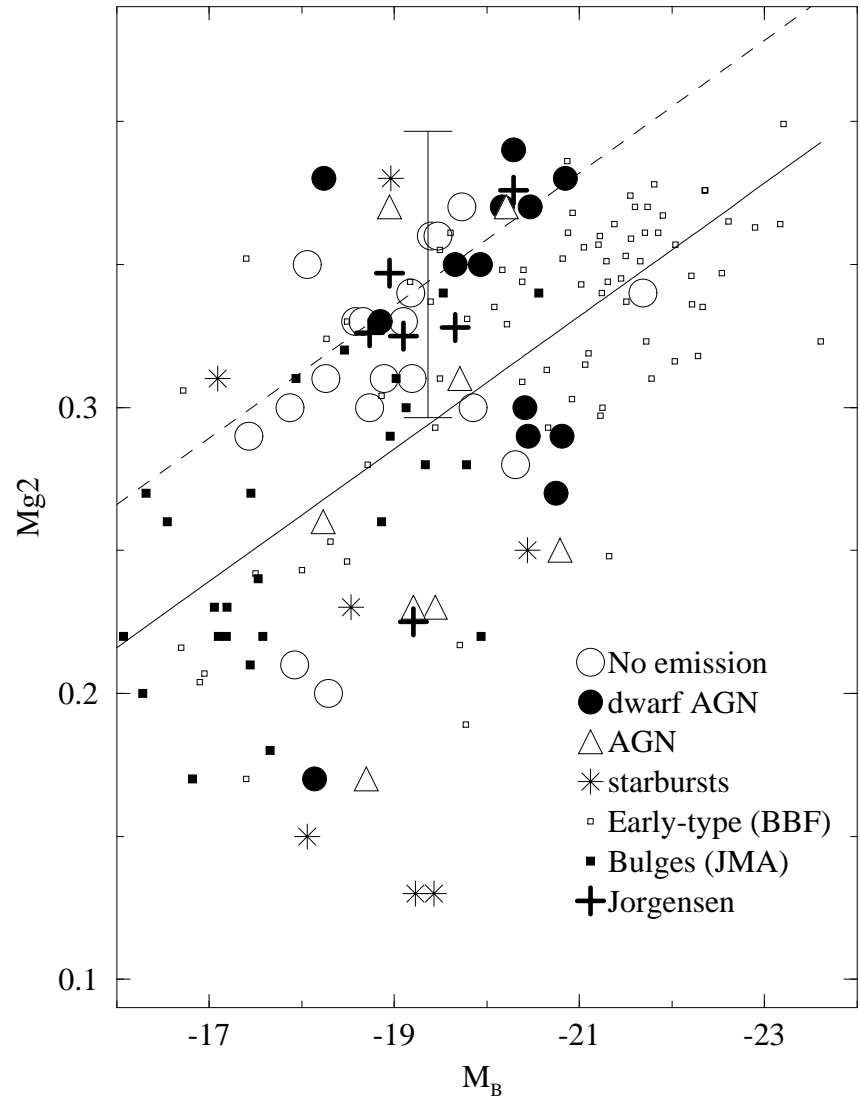
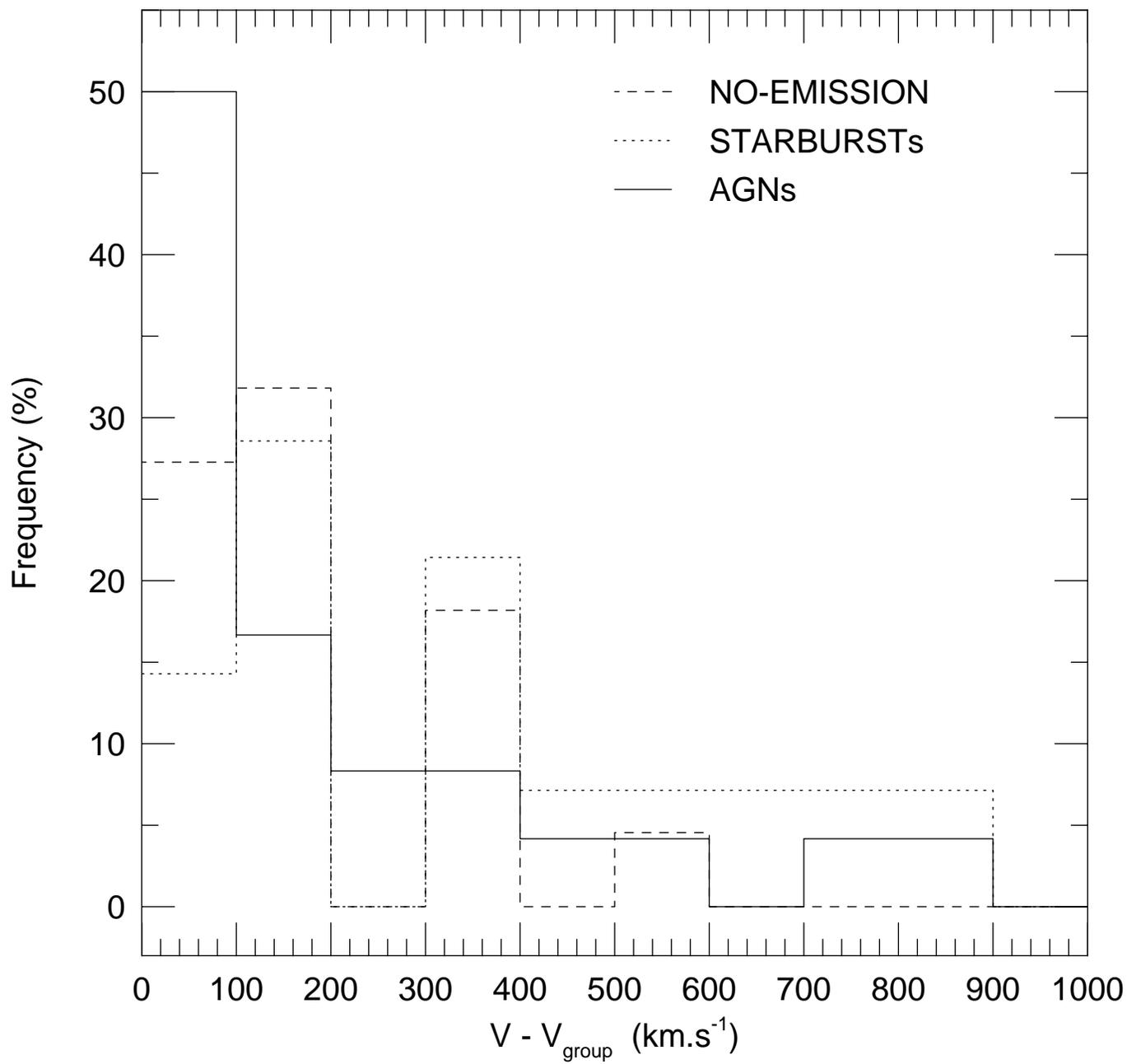
\end{document}